\documentclass[reqno,a4paper]{article}
\usepackage{coling2016}
\usepackage[markers,figuresonly,nofiglist]{endfloat}
\usepackage{amssymb,amsmath,amsthm}
\usepackage[T1]{fontenc}

\title{\Huge{\textbf{Is the PPG signal chaotic?}}}
\author{Javier de Pedro-Carracedo\thanks{Current address: Computer Engineering Department, University of Alcal\'a (UAH), E-28871 Alcal\'a de Henares, Spain.} \\
  Photonic Technology and Bioengineering Department\\
  Technical University of Madrid (UPM)\\
  E-28040 Madrid, Spain \\
  \texttt{javier.depedro@uah.es} \\
   \AND
 David Fuentes-Jimenez \\
  Department of Electronics \\
  University of Alcal\'a (UAH) \\
  E-28871 Alcal\'a de Henares, Spain \\
  \texttt{david.fuentes@depeca.uah.es} \\
  \AND
  Ana M. Ugena \\
  Departamento de Matem\'atica aplicada a las Tecnolog\'ias de la Informaci\'on \\
  Technical University of Madrid (UPM) \\
  E-28040 Madrid, Spain \\
  \texttt{anamaria.ugena@upm.es} \\
  \AND
  Ana P. Gonzalez-Marcos \\
  Photonic Technology and Bioengineering Department \\
  Technical University of Madrid (UPM) \\
  E-28040 Madrid, Spain \\
  \texttt{anapilar.gonzalez@upm.es} \\
}

\date{}
\theoremstyle{remark}

\usepackage{amsmath}
\usepackage{amssymb}
\usepackage{multirow}
\usepackage{float}
\usepackage{subfig}
\usepackage{makecell}
\usepackage{adjustbox}
\usepackage{textcomp}

\usepackage{cleveref}
\Crefname{figure}{Figure}{Figures}

\usepackage{booktabs,dcolumn}
\DeclareMathVersion{nxbold}
\SetSymbolFont{operators}{nxbold}{OT1}{cmr} {b}{n}
\SetSymbolFont{letters}  {nxbold}{OML}{cmm} {b}{it}
\SetSymbolFont{symbols}  {nxbold}{OMS}{cmsy}{b}{n}
\newcolumntype{L}{D{.}{.}{-1}}

\makeatletter
\newcolumntype{B}[3]{>{\boldmath\DC@{#1}{#2}{#3}}c<{\DC@end}}
\newcolumntype{Z}[3]{>{\mathversion{nxbold}\DC@{#1}{#2}{#3}}c<{\DC@end}}
\makeatother

\begin{document}

\maketitle

\begin{abstract}
\noindent This paper shows how the dynamics of the PhotoPlethysmoGraphic (PPG) signal, an easily accessible biological signal from which valuable diagnostic information can be extracted, of young and healthy individuals performs at different timescales. On a small timescale, the dynamic behavior of the PPG signal is predominantly quasi-periodic. On a large timescale, a more complex dynamic diversity emerges, but never a chaotic behavior as earlier studies had reported. The procedure by which the dynamics of the PPG signal is determined consists of contrasting the dynamics of a PPG signal with well-known dynamics---named reference signals in this study---, mostly present in physical systems, such as periodic, quasi-periodic, aperiodic, chaotic or random dynamics. For this purpose, this paper provides two methods of analysis based on Deep Neural Network (DNN) architectures. The former uses a Convolutional Neural Network (CNN) architecture model. Upon training with reference signals, the CNN model identifies the dynamics present in the PPG signal at different timescales, assigning, according to a classification process, an occurrence probability to each of them. The latter uses a Recurrent Neural Network (RNN) based on a Long Short-Term Memory (LSTM) architecture. With each of the signals, whether reference signals or PPG signals, the RNN model infers an evolution function (nonlinear regression model) based on training data, and considers its predictive capability over a relatively short time horizon.
\end{abstract}

\medskip


\section{Introduction}\label{sec:IN}

Even in ancient times, medical specialists paid particular attention to human physiology, since understanding the mechanisms that make it possible for the human body to function correctly opened up a new avenue in development diagnostic procedures to possible pathologies or more and less severe somatic disorders \cite{Nguyen2016}. Physiological systems are dissipative systems that, in their energy exchange with the surrounding environment, ensure the stability of the homeostatic process, an internal self-regulatory mechanism that guarantees optimal vital conditions \cite{Haddad2005}. This interrelation is best reflected by dynamic variables that can be measured directly or indirectly using the appropriate technical equipment. Dynamic variables are known in medical jargon as biological signals and represent the dynamic response (output variables) of the physiological system at hand. Roughly speaking, it is not always possible to measure all the dynamic variables involved in physiological functioning; at best, only a few, the so-called physical observables, are available. Even then, it is practically possible to gain information on the state of the system only using one dynamic variable \cite{Packard1980}.

The self-regulatory gearing of physiological systems operates intricately at different timescales, even though, in many cases, the time response of some of the system's dynamic variables shows an apparent regularity at small timescales \cite{Dana2009}. However, this apparent regularity conceals dynamic subtleties that mainly expect materialize in the longer term and which often mask under the guise of dynamical noise. These minor variations provide the necessary complexity to enable regulation at a very precise and effective level. Therefore, biological signals have both deterministic and stochastic components \cite{Costa2005}, both contributing to the underlying dynamics of the physiological system. 

In a previous paper, we study the dynamic behavior of the PPG signal from a stochastic perspective. Through a slight modification of the 0--1 test, we contrast the dynamic behavior of the PPG signal with a diffusive process at different timescales. The results compare with those obtained for reference signals whose dynamic behavior is known, aside from being typical of the possible dynamics that physical systems may exhibit \cite{Pedro-Carracedo2019}. This paper assumes no a priori analytical approach. This study intends to inquire, using neural network-based methods, which dynamics are present in PPG signals of young healthy individuals with the known dynamics of the reference signals. By learning these known dynamics, even in the presence of noise, together with their predictive capability, the neural networks in this work have made it possible to examine the dynamic composition imbedded in the time evolution of any PPG signal at different timescales.  

The rest of the paper is organized as follows. Section \ref{sec:MAM} describes the datasets used in the experimental tests, as well as the two deep neural network-based architectures that serve as an experimental framework. In addition to the adapted internal structure of these architectures, this section refers to their training to fulfill analytical requirements pursued in this study. Section \ref{sec:Results} shows the obtained results, both graphically and numerically, for various experimental settings. In section \ref{sec:Discussion}, we analyze and interpret the obtained results. Finally, in section \ref{sec:Conclusions}, we briefly outline the conclusions drawn from this study, which in turn serve as the basis for future work.

\begin{figure*}[ht!]
\centering
\subfloat[]{\includegraphics[scale=0.46]{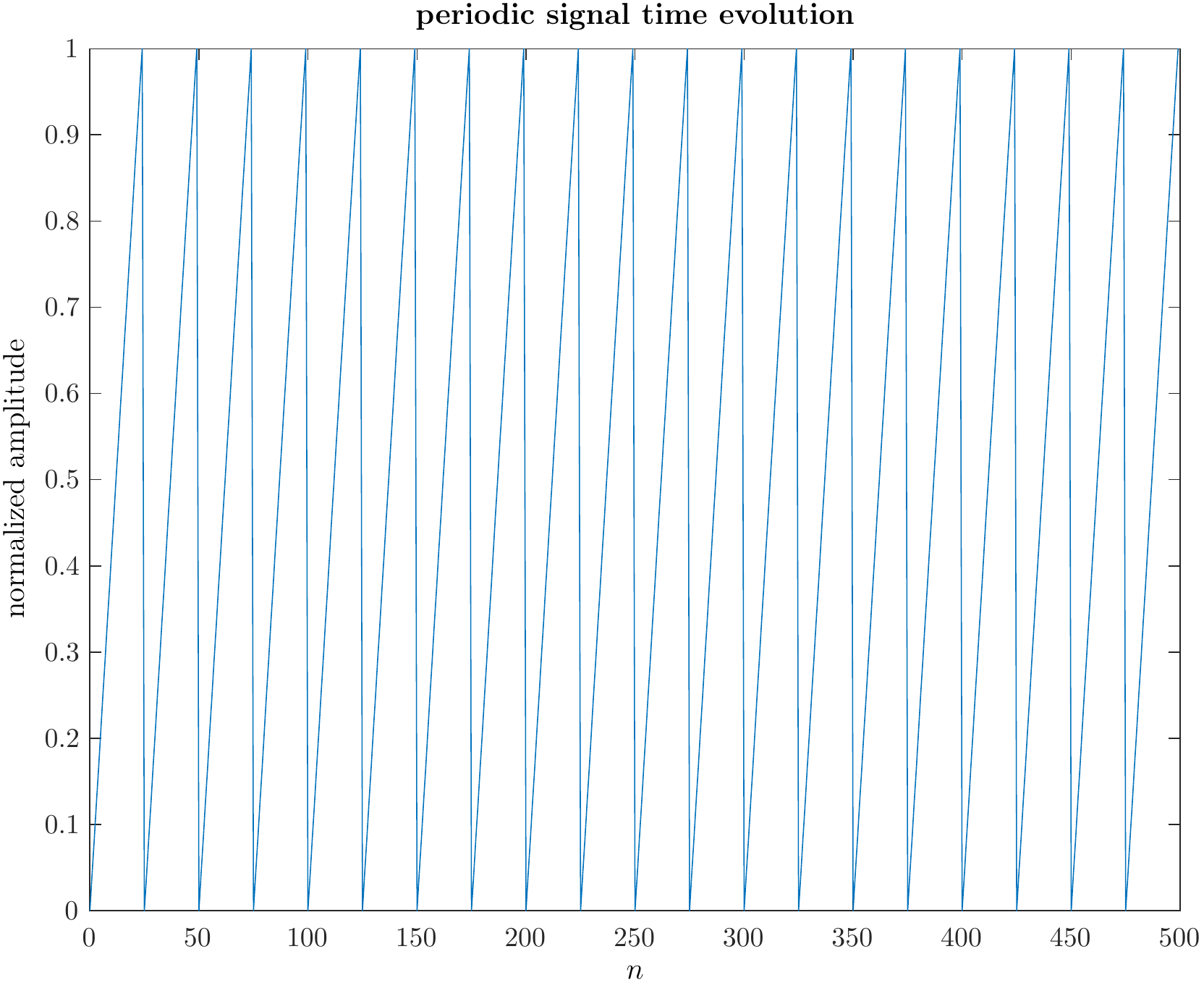}
\label{fig:FRSTEa}}
\subfloat[]{\includegraphics[scale=0.46]{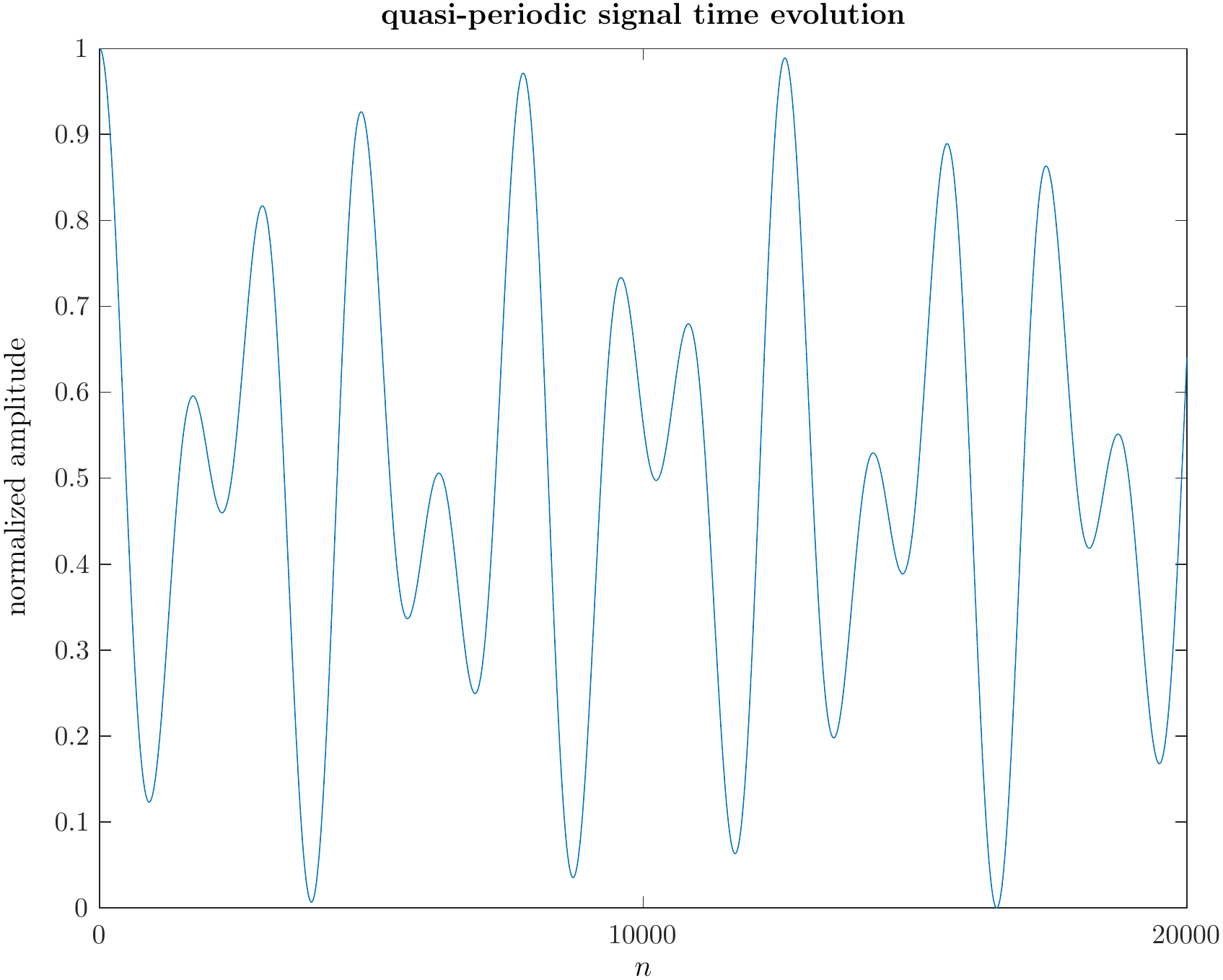}
\label{fig:FRSTEb}}\hfil
\subfloat[]{\includegraphics[scale=0.46]{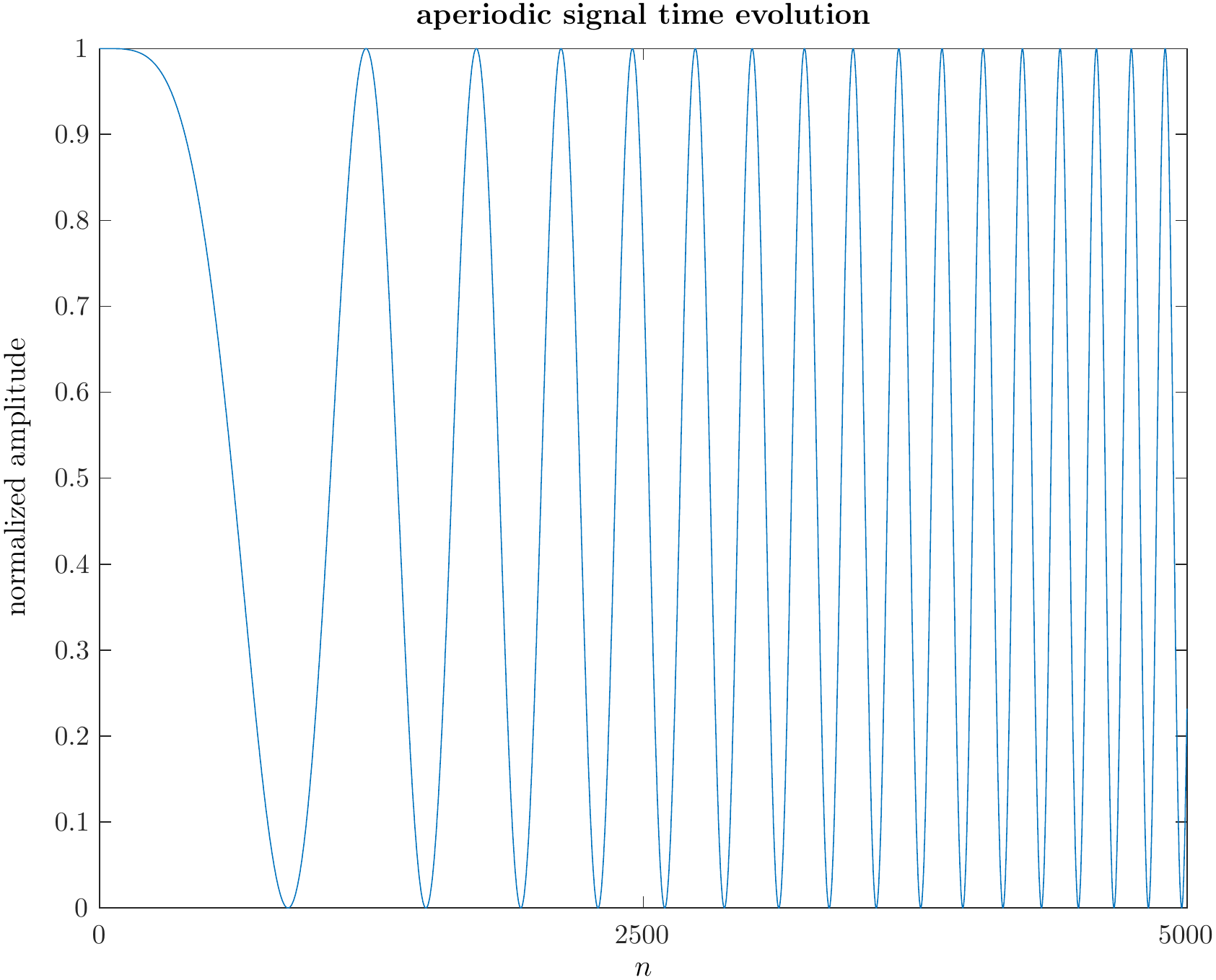}
\label{fig:FRSTEc}}
\subfloat[]{\includegraphics[scale=0.46]{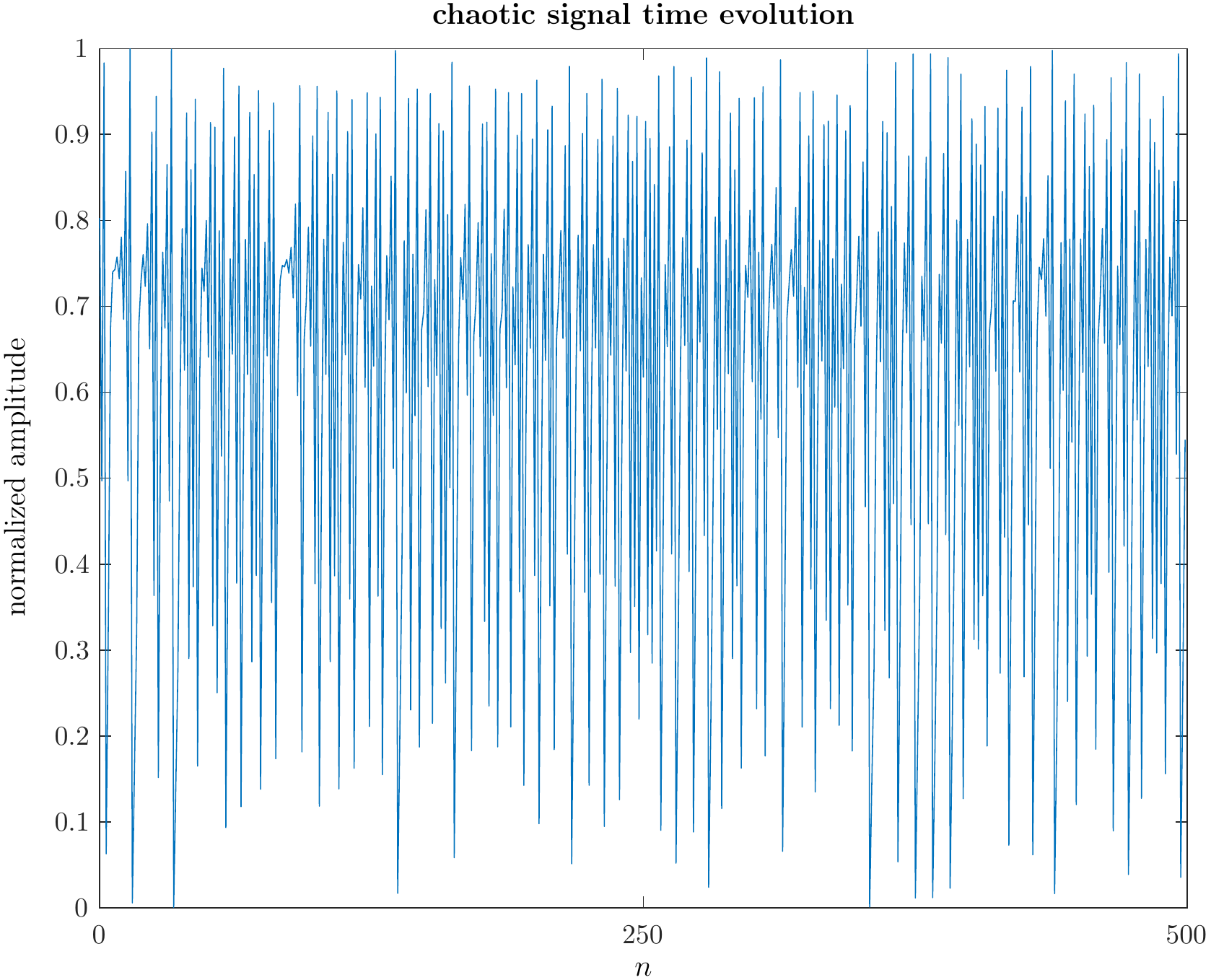}
\label{fig:FRSTEd}}\hfil
\subfloat[]{\includegraphics[scale=0.46]{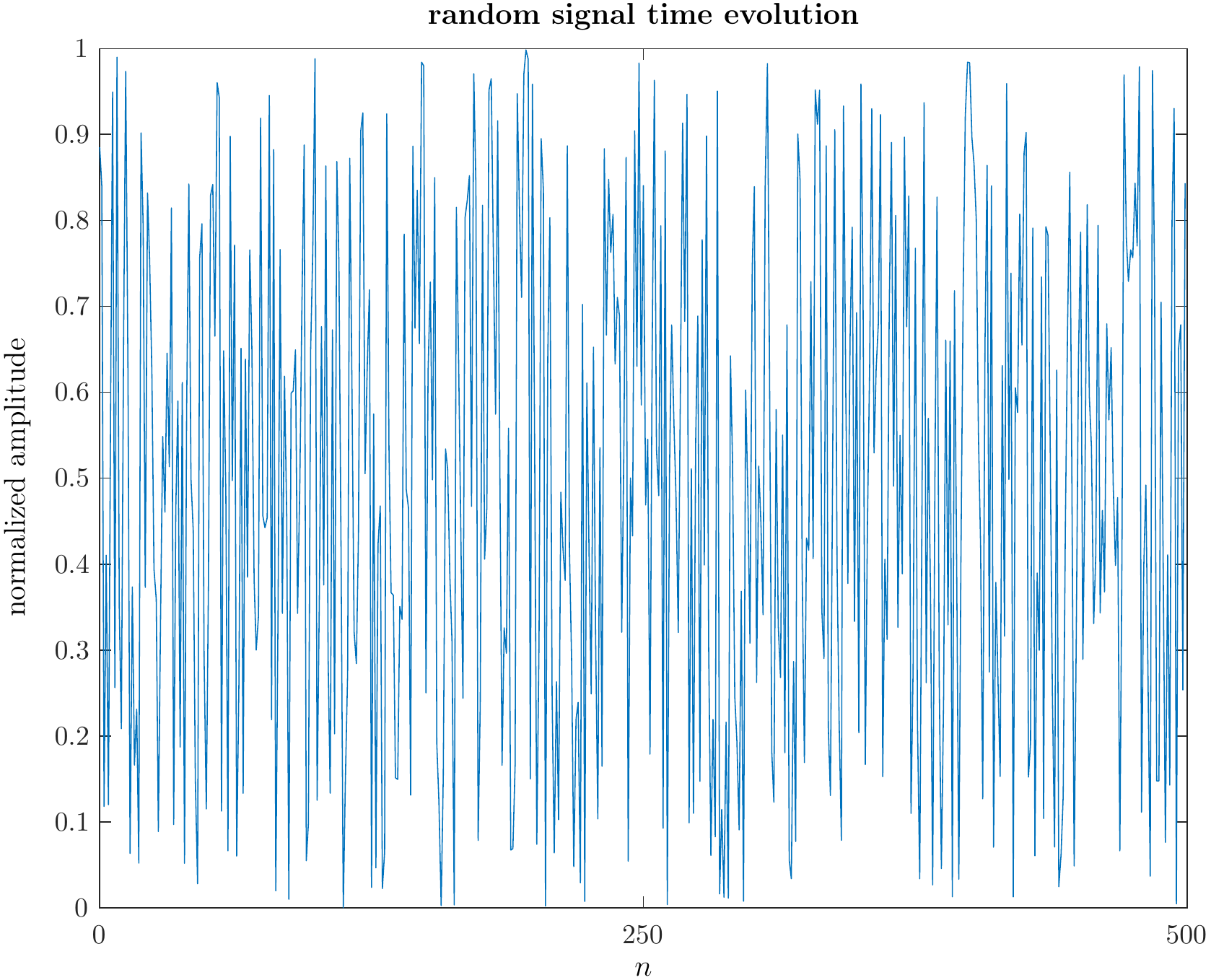}
\label{fig:FRSTEe}}
\caption{Synthetically generated training signals (reference signals). For visual clarity in each signal the interval that best identifies its time structure is chosen: (\textbf{a}) Periodic signal; (\textbf{b}) Quasi-periodic signal; (\textbf{c}) Aperiodic signal; (\textbf{d}) Chaotic signal; (\textbf{e}) Random signal.}
\label{fig:FRSTE}
\end{figure*}

\section{Materials and Methods}\label{sec:MAM}
\subsection{Datasets}\label{ssec:DATA}
\subsubsection{Reference signals (synthetic dataset)}\label{sssec:syndata}

Although the complexity of signals admits different definitions---a rigorous definition has not been agreed by the scientific community \cite{Grassberger1986,Badii2003}---, a consistent alternative evaluates the degree of complexity regarding regularity in the patterns of repetition of the data; to some extent, this alternative quantifies the difficulty in describing or understanding a signal \cite{kantz2004}. These patterns lead to an ordering of the signals between two opposite ends, the most regular or periodic and the most irregular or random, with a whole range of intermediate options; as they approach a random regime, as in the case of chaotic signals, these acquire greater degrees of freedom or versatility, without losing determinism in its dynamic behavior.

To train our CNNs, we have used 150,000 samples (points) of each one of the different signals that represent the typical behavior of time series concerning regularity in the repeating patterns of the data (reference signals). Classification of the signals is according to the order of less to greater complexity, in accordance with the aforementioned approach, from the most regular or periodic evolution to the most random one. Figure \ref{fig:FRSTE} shows the amplitude variation, normalized to the interval $[0, 1]$, of all reference signals in the time domain. 

\paragraph{Periodic signal}

Here, we analyze a repetitive pattern with two frequencies whose values are rationally related, i.e., we use a saw's wave at $f=10$ Hz:

\begin{equation}\label{eqn:SP}
y_{\text{periodic}}(t)=2\left(\frac{t}{1/f}-\left\lfloor \frac{1}{2}+\frac{t}{1/f}\right\rfloor\right).
\end{equation}

\paragraph{Quasi-periodic signal}

A certain recurrence in time evolution may lead to wrongly consider this type of dynamics as an ``irregular periodicity'' or the same ``repetitive structure''. The fact is that the pattern never repeats if the data have infinite precision, shaping a torus or tori depending on the degrees of freedom \cite{Landau1987}. In our case, we apply the most simple one, two cosine waves whose frequencies are irrationally related:

\begin{equation}\label{eqn:SQ}
y_{\text{quasi-periodic}}(t)=\cos\left(1\cdot t\right)+\cos\left(\omega_{1}\cdot t \right),
\end{equation}
with $\omega_{1}$ being the reciprocal of the golden mean.

\paragraph{Aperiodic signal} 

An aperiodic signal has non self-similar repetition even with infinite precision data, although mathematically it can be considered like a periodic function with an infinite period. The selected function is the one that generates samples by a linear frequency sweeping, the \textit{chirp} function: $f(t)=f_{0}+k\cdot t$, between $f_{0}=0$ Hz and $f_{1}=10$ Hz, with $k=(f_{1}-f_{0})/T$, being $T$ the sweep time; and the sampling frequency $f_{s}=250$ Hz.

\begin{equation}\label{eqn:SA}
y_{\text{aperiodic}}(t)=\sin\left[2\pi\left(f_{0}t+\frac{k}{2}t^{2}\right)\right].
\end{equation}

\paragraph{Chaotic signal}

The sensitivity of certain deterministic functions to small changes in the initial state is the characteristic footprint of chaotic behavior. The initial uncertainty increases with time, and it is not possible to predict the final state of the system (N. S. Krylov 1944; M. Born 1952) \cite{Landau1987}. We use the H\'enon map dynamic variable $x_{n}$, with $a=1.4$ y $b=0.3$. The initial conditions were $x_ {0}= y_{0}= 0.03$. One million samples were generated, using the last 150,000 points. 

\begin{eqnarray}\label{eqn:SC}
x_{n+1} &=& 1-ax_{n}^{2}+y_{n},\nonumber \\
y_{n+1} &=& bx_{n}.
\end{eqnarray}

\paragraph{Random signal} 

It is not deterministic and requires a probabilistic characterization. 150,000 points were generated by a uniform distribution in the interval $[0,1]$.

\subsubsection{Real-world PPG signals (biological dataset)}\label{sssec:realdata}

This paper focuses only on a single biological signal, the PhotoPlethysmoGraphic (PPG) signal; forward publications will describe the results for more biological signals. We have chosen the PPG signal because it is easily accessible and the information provided allows us to monitor vital physiological signs. A pulse oximeter consists of a light emitter and a photodetector that collects and records (pulse or PPG signal) the loss---scattering and absorption---that a beam of light undergoes when it passes or reflects from human tissue. It allows detecting blood volume changes in the microvascular bed of tissue---in our case, the middle finger of the left hand---, obtaining valuable information about the cardiovascular system and, on the whole, about the cardiorespiratory system. Given the simplicity of its non-invasive accommodation, in addition to its low cost, a pulse oximeter is very useful in biomedical applications for clinical and sports environments. For a useful review refer to \cite{Allen2007}.

In this work, we use the PPG signals from a total of 40 students, between 18 and 30 years old and a non-regular consumers of psychotropic substances, alcohol or tobacco, selected to participate in a national research study \cite{Aguilo2015,Arza2018}. All signals were captured from the middle finger of the left hand and sampled at a frequency of 250 Hz \cite{Aguilo2015}, say, sampling time $\Delta t=4$ ms. Since the PPG signal operates at a very low frequency, where its dynamic diversity is manifested, it is necessary to work with many points; hence we have used up to 600,000 points (40 minutes) of every PPG signal, those corresponding to the total data recorded in the indicated research project.

\subsection{Dynamic behavior classification with a CNN architecture}\label{ssec:CNNarch}

In this paper, we propose a CNN-based approach to classify PPG signals according to their dynamic behavior. The CNN receives time segments of the PPG signal as input, and it produces a normalized matching-up vector (\textit{target pattern}) as output that contemplates the success rate of each possible dynamic behavior assignable to the input signal. The available dynamic behaviors are periodic, quasi-periodic, aperiodic, random and chaotic. 

\subsubsection{CNN architecture}

Here we propose an encoder-fully connected residual architecture whose residual blocks are based in the ones from \cite{He2016}. The encoder-fully connected architectures are traditionally used in classification tasks like  \cite{Simonyan2015,Szegedy2015}, actually it is used in a range of applications like Wi-Fi people detection \cite{Huang2020} or shape regression \cite{Guler2017}. We define the proposed CNN as ours for three main reasons: first, that is a CNN explicitly created for 1D signal processing; at next, we prepare an structure capable of process high and low frequencies with the use of long kernels; finally, we base our residual blocks in ResNet \cite{He2016}, but all of them have been adapted to 1D processing. 

\begin{figure}[ht!]
\centering
\includegraphics[width=0.9\columnwidth]{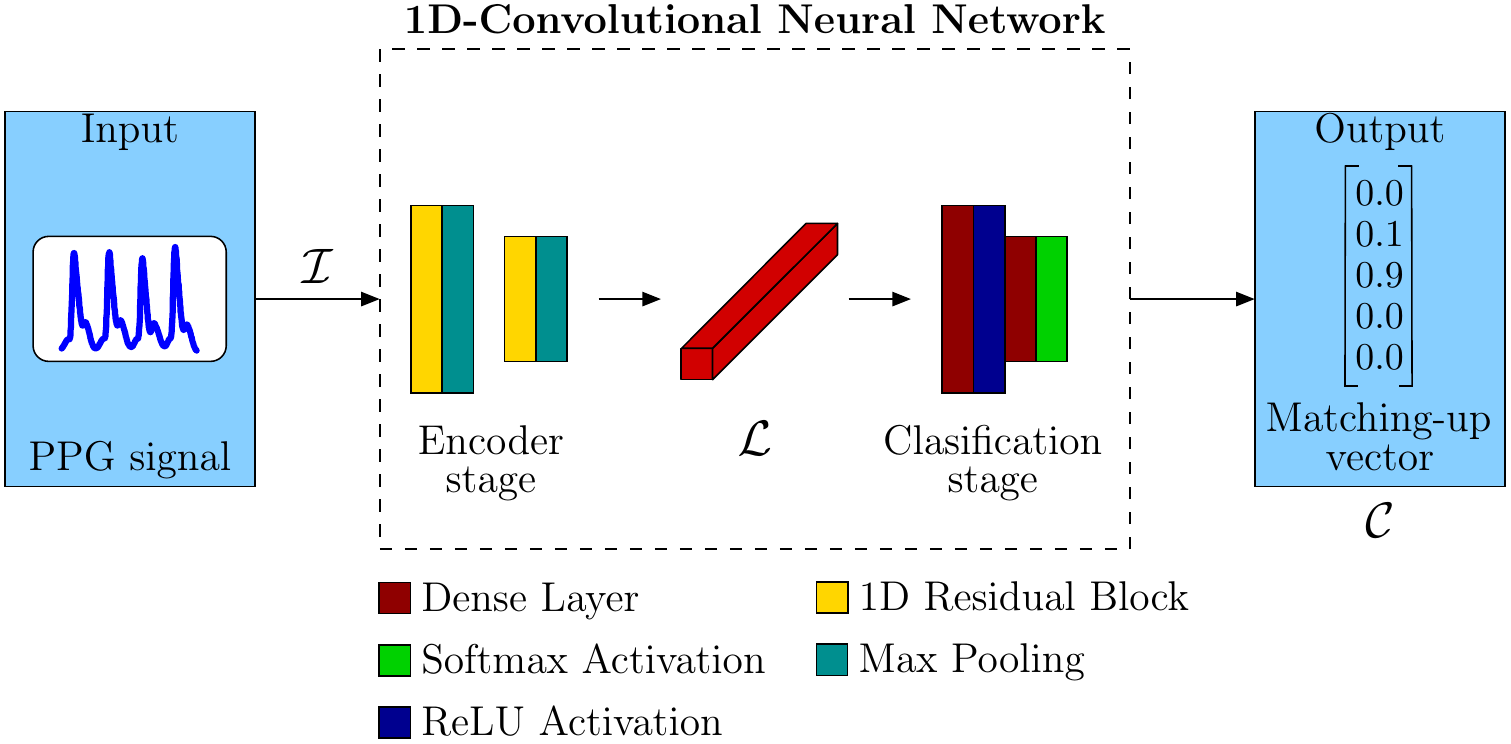}
\caption{1D CNN proposed architecture.}\label{fig:FCNNdiag}
\end{figure}

 The reason to use residual block configurations is that it allows preserving all the processed information, a very known problem in CNN architectures. With these residual blocks, we can process and add new information to the old one, enriching the information to be processed by the next layer, in contrast to non-residual structures like the ones used in \cite{Simonyan2015}, that loss information across each one of the layers used. In our case, both low and high-frequency information is essential (specially low frequency), for which reason is very important to preserve or reuse all the available information. We cannot afford to lose it, because all this information can hide invaluable dynamics behaviors in the PPG signals.

The base structure is shaped like a classic 1D-CNN that has two stages. The first one is an encoder feature extractor that is composed of 1D residual blocks whose basis is the one used in \cite{He2016}. This stage includes max poolings layers that obtain the most important features and erase the spurious of the previous outputs. The second one is a classification stage composed by fully connected layers, where the features obtained by the encoder are vectorized and used as inputs to the fully connected layers, that are the ones responsible for classifying the dynamics behavior of the input signal segment using the encoder output features that characterize the input signals.

The input time segment $\mathcal{I}$ is processed by the encoder feature extractor, which generates the latent filtered representation $\mathcal{L}$. The process from the input signal to obtaining the latent space follows these steps. Firstly, the point segment of the input signal passes through the first residual block used; this first residual block has a kernel of 14 so that its receptive field will be quite wide, being able to capture both high and low frequencies. This first residual block is in charge of obtaining the simplest filters characterized by simple features such as fast transitions between samples or simple relationships between them. In order to filter the outputs of this residual block, a max pooling layer is used, which allows eliminating parasitic activations and highlighting the most outstanding ones. Once the first residual block $+$ max pooling has been passed, the second residual block is used. This second block, unlike the first one, uses a lower kernel since it does not require such a wide receptive field when using as input the outputs of the filters of the first residual block. The mission of this second block is to obtain more complex features and of greater dimensionality; once again, a max pooling layer is used at the output to highlight which of these more complex features are best suited to the input signal segment. This last output is the one that constitutes the latent space of our neural network. Once we generate the latent space, we proceed to the classification stage or fully connected. In this stage, the first thing we do is to vectorize the input so that the dense layers can easily process it. After vectorizing the latent space, it passes through a first dense layer composed of 12 neurons and a ReLU activation; this is the first one to process the information before its final classification. The second and final stage is a 5-neuron dense layer, which corresponds to the number of output dynamic behaviors. This last layer has a softmax activation that allows obtaining some output dynamics with a success rate and normalized format to one; besides, it is the one in charge to relate finally all the information used and to obtain the final criterion of classification of the dynamics. So this final stage produces the 5-dimensional output vector $\mathcal{C}$, as we can see in Figure \ref{fig:FCNNdiag}.

\setcellgapes{5pt}\makegapedcells
\begin{table}[ht!]
\caption{1D Proposed CNN detailed architecture.}\label{tab:TCNNarch}
\centering
\begin{adjustbox}{max width=\linewidth}
\begin{tabular}{|c|c|c|c|}
\hline
\textbf{Layer number} & \textbf{Type}  & \textbf{Output size} & \textbf{Parameters} \\ 
\hline
$1$ & Input & $(5000,1)$ & --- \\ 
\hline
$2$ & Residual Block  & $(4870,32)$  & $\text{Kernel}=14/\text{Activation}=\text{ReLU}$ \\
\hline
$3$ & Max Pooling 1D & $(2420,32)$ &  $\text{Pool}=5$  \\
\hline
$4$ & Residual Block & $(2360,64)$ & $\text{Kernel}=7/\text{Activation}=\text{ReLU}$ \\	
\hline
$5$ & Max Pooling 1D & $(1160,64)$ & $\text{Pool}=5$ \\
\hline
$6$ & Flatten & $(74240)$ & --- \\		
\hline
$7$ & Fully Connected & $(12)$ & $\text{Activation}=\text{ReLU}$ \\	
\hline			
$8$ & Fully Connected & $(5)$ & $\text{Activation}=\text{Softmax}$ \\	
\hline
\end{tabular}
\end{adjustbox}
\end{table}

\begin{figure}[ht!]
\centering
\includegraphics[width=0.5\columnwidth]{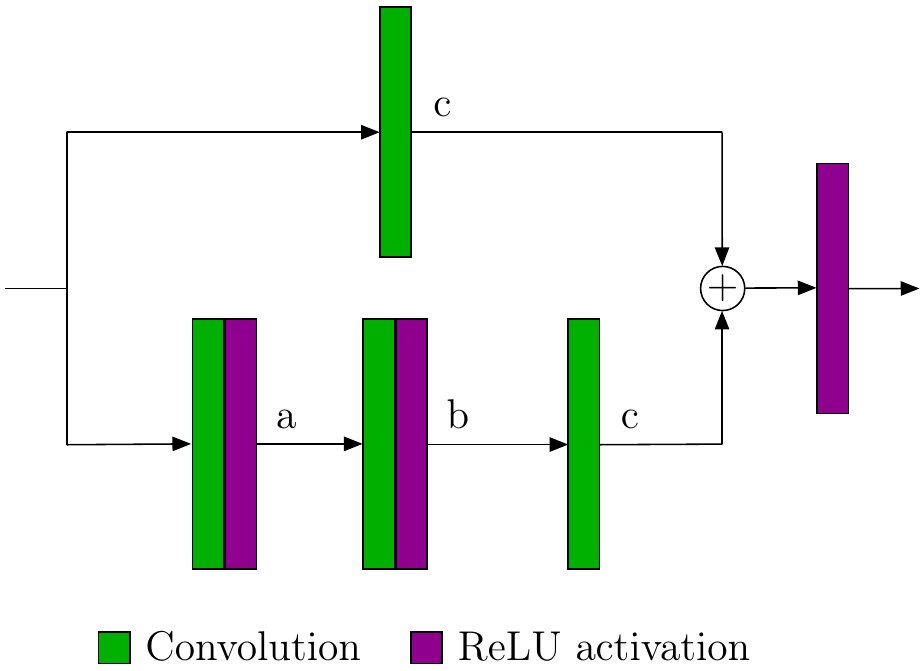}
\caption{Residual blocks, based on \cite{He2016}. These blocks have two parallel branches, in which the first one has a single convolution. In contrast, the second uses three convolutions followed by ReLU activations that first reduce the dimensionality of the input through the number of filters and then increase it in the second convolution creating more complex filters and finally in the last convolution project it through a unitary convolution in a space of lower dimensionality. This structure allows the use of convolutions at different scales, combining their outputs without losing much information and acting in terms of size as a traditional convolutional layer.}\label{fig:FCNNblocks}
\end{figure}

In Table \ref{tab:TCNNarch}, we can see the architecture in detail, with all its layers and parameters. Finally, to provide a better understanding of the used residual blocks insides, we graphically display all its structure in Figure \ref{fig:FCNNblocks}.

In short, the reason to propose and use this architecture is given by its very favorable characteristics to face the problem here posed. Such characteristics, in summary, are composed of:  

\begin{itemize}
\item \textbf{Wide Kernels:} These kernels allow a receptive field captures both high and lower frequencies, being able to unravel all the dynamic behaviors present in the signal at different timescales.
\item \textbf{Residual Blocks:} These blocks allow, on the one hand, improve the convergence of training and, on the other hand, allow reuse all the information that is lost through the convolutional filters, reducing the information lost to a minimum.
\item \textbf{Max Pooling Layers:} These layers allow filtering the most negligible and noisy components generated through the neural network.
\end{itemize}

 To test the effectiveness of our proposed architecture, a comparison is made in \ref{sssec:comparison}.

\subsubsection{Training}

In the training step, we use the reference signals explained previously in \ref{sssec:syndata} to train our CNN end to end and allowing this to learn the most important features of the dynamic behavior matching task, allowing us to avoid the use of a real-world signal labeled database. In this step, we train the network with the synthetic dataset, which is divided into training, validation and testing sets, which are composed of 80\%, 10\% and 10\% of each dataset. These division ranges have been chosen since usually the training set constitutes 60 to 80 per cent of the dataset while the validation and test set comprises 20 to 40 per cent. At next, we explain the training parameters.

\begin{enumerate}
\item \textbf{Optimizer}. The used optimizer is the Adaptative Moment Estimation, better known as \textit{Adam} \cite{Kingma2014}. This optimizer is an alternative to the traditional stochastic gradient descent (SGD) algorithm optimizer, and it combines the advantages of two previous alternatives \cite{Duchi2011,Dauphin2015}, creating a more reliable preference that uses the averages of the first and second moments of the gradient to adapt the learning rate dynamically. The learning rate is the parameter that defines how much and how fast our system learns in each period; a very high learning rate may result in training divergence, while a very low learning rate may not advance or take a long time to train the system. \textit{Adam} starts with a user-defined learning rate and, after that, it modifies the learning rate along with the training without supervision, providing an adaptable learning rate to the train, which is a big advantage in terms of learning adaptation. The initial learning rate is $10^{-4}$; to complement this train, we use an early stopping callback. The early stopping callback is a tool that helps us to save the best model of all our training. This callback records the metrics and losses achieved in each epoch, saving only the best of the achieved models. The network is trained for 10 epochs with a batch size of 50 samples, but the before mentioned callback saves the best-obtained model in these 10 epochs. The total training time is about 6 hours.
\item \textbf{Loss Function}. Our CNN uses to train a training set formed by input signal segment $\mathcal{I}$ and output matching-up vector $\mathcal{C}$. The proposed multiclass classification loss function is the categorical cross-entropy loss, that evaluates the differences between ground truth and predictions. The categorical cross-entropy loss function is applied between the ground truth and the per class calculated success rate after softmax activation by the CNN $S_{j}$. In Eq. \eqref{eqn:E01cross}, $S_{p}$ appears when using \textit{one hot} representation for ground truth vectors, so that there's only one non-zero element $t_{p}=1$, which belongs to the true label.

\begin{equation}\label{eqn:E01cross}
\text{CCE}=-\log \left(S_{p}\right).
\end{equation}

\end{enumerate}

Finally, we explain in-depth and step by step how the training takes place. The first thing that needs to be clarified is that it is necessary to train one CNN for the small timescale, whose inputs are represented by signal segments of 5,000 samples and another one that is in charge of the large timescale whose inputs are composed of signal segments of 60,000 samples. The training of these CNNs is done through the synthesized reference signals, which are composed of periodic, quasi-periodic, aperiodic, random and chaotic dynamics. Since these CNNs have been designed from a classification point of view, the problem to be solved is a multi-class classification, with five possible classes or dynamic behaviors to be found. Each of the segments of the input signals has associated output labels defined, which directly link these input segments to a specific dynamic, learning the system that with $100\%$ success rate, these segments are associated with a dynamic behavior. In the training process, a random batch generator is used that takes 50 random segments from the 5 reference signals with 5,000 or 60,000 points and generates 50 output labels assigned. This process has been randomized to provide more generalized and learning-rich training. Once the batches are loaded, during 10 periods and through the use of the \textit{Adam} optimizer, the proposed system is trained to recognize the input signals correctly. For the training to be carried out, an $80\%$ (600,000 points) of data is used for training, a $10\%$ (75,000 points) of data for validation, and a $10\%$ (75,000 points) of data for test. After training the proposed system with the reference signals, it is evaluated with the real-worlds PPG signals, which, unlike the reference signals, have a dynamic with multiple dynamic behaviors at different timescales. So, its output vector does not have a single dynamic with a $100\%$ success rate, but this rate is redistributed among the five dynamics trained through the reference signals, obtaining outputs characterized by a composite dynamics and not unique.

\subsection{Horizon of prediction with an RNN architecture}\label{ssec:RNNarch}

We now propose an RNN architecture to predict the following $h_{p}$ points of the time series conformed by the PPG signals beyond the current point considered in a current time $t_{c}$. In addition to this, we predict the following $h_{p}$ points of periodic, quasi-periodic, aperiodic, random and chaotic time series, to offer a perspective of how predictable are the PPG signals and with which type of dynamic horizon or horizons are most associated. The RNN receives time segments of the PPG signal as input and it outputs a regression of the $t_{c}+h_{p(n)}$ point, being $p(n)$ the offset regarding to the current time (starting point).

The horizon of prediction $h_{p}$ is 50 points to cover a wide range of samples to stabilize the prediction error. Thus, we train 50 different RNNs for each time series specialized in predict the next 50 points with regards to the starting point.

\subsubsection{Architecture}

 We suggest the use of an RNN. The best-known types of RNN are, firstly, the classic RNN blocks, which are simply made up of a tanh activation and a concatenation of the current input with the previous outputs. These blocks are in disuse due to their low efficiency and high instability, as they diverge significantly. Secondly, the LSTM or Long-Short Term Memories units introduced the concept of doors in recurrent networks, using different structures within the same block to carry out different functions (door of oblivion, update door, exit door), each one vital for the correct operation of the same. Finally, the Gate Recurrent Units (GRU) \cite{Cho2014}, which are a simplification of the LSTM, reducing the number of structures or gates to two, the reset gate and the update gate, would theoretically obtain similar performances.
 
\begin{figure}[ht!]
\centering
\includegraphics[width=0.9\columnwidth]{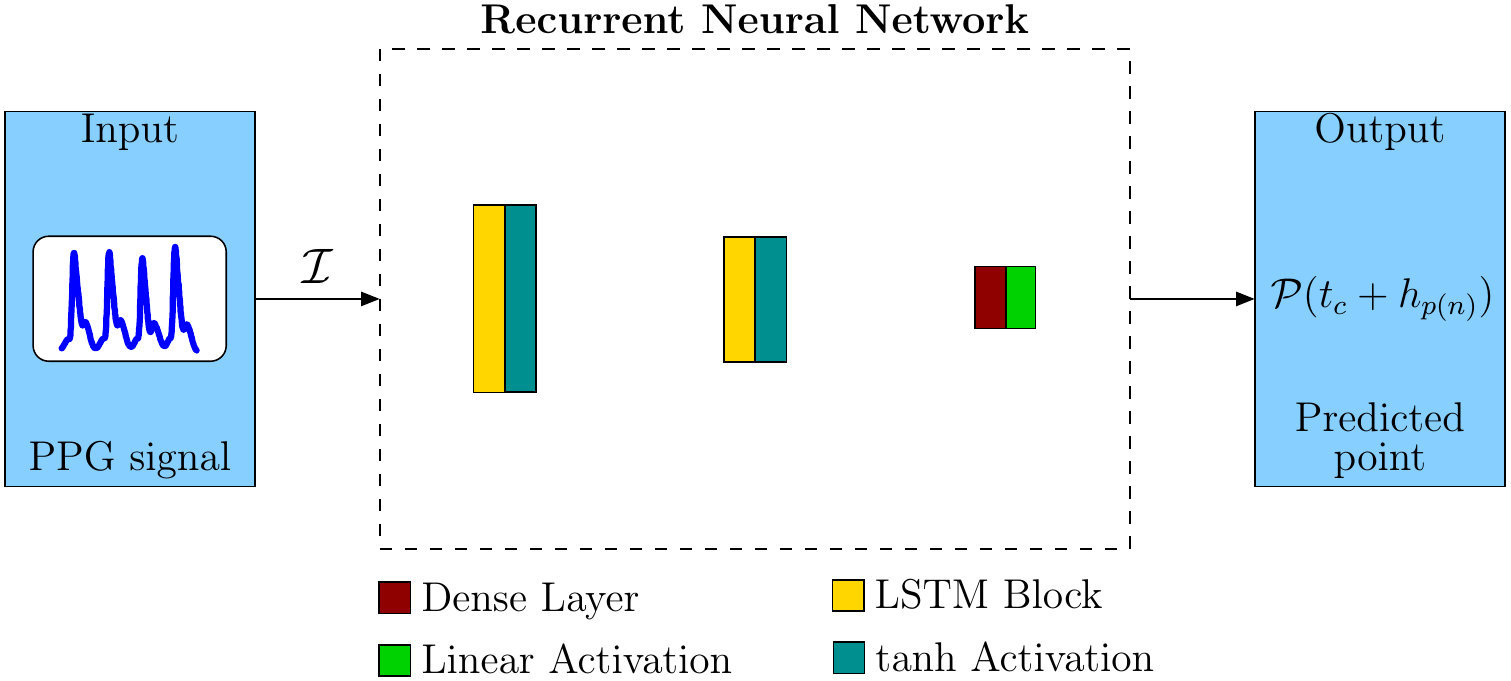}
\caption{Proposed RNN architecture.}\label{fig:FRNNdiag}
\end{figure}
 
Here we propose a Recurrent Neural Network based in LSTM structures \cite{Hochreiter1997}, since the RNN are highly unstable and problematic, while the GRU are recurrent units that try to simplify the LSTM units and that theoretically obtain the same performance as these, even though this adjustment may not have been observed in practice. The LSTM-based architectures are traditionally used in time series regression and classification tasks, as we can see in \cite{Zhang2018,Ningsih2019,Khotimah2019}.

\setcellgapes{5pt}\makegapedcells
\begin{table}[ht!]
\caption{Proposed RNN detailed architecture.}\label{tab:TDNNarch}
\centering
\begin{adjustbox}{max width=\linewidth}
\begin{tabular}{|c|c|c|c|}
\hline
\textbf{Layer number} & \textbf{Type}  & \textbf{Output size} & \textbf{Parameters} \\ 
\hline
$1$ & Input & $(1,1200)$ & --- \\
\hline
$2$ & LSTM  & $(100)$  & $\text{Number of Units}=100/\text{Activation}=\text{Hyperbolic Tangent}$ \\
\hline
$3$ & LSTM  & $(20)$  & $\text{Number of Units}=20/\text{Activation}=\text{Hyperbolic Tangent}$ \\
\hline
$4$ & Fully Connected & $(1)$ & $\text{Activation}=\text{Linear}$  \\		
\hline
\end{tabular}
\end{adjustbox}
\end{table}

Two LSTM layers with sequence return connections make up the base structure and one final dense layer to provide the point prediction. The first layer LSTM is in charge of obtaining the time characteristics of the input signals directly. This first layer returns as output the final state and all those intermediate states of the LSTM, which allow enriching the output of this layer widely. This enrichment is used by the second layer LSTM that uses all these time characteristics and obtains much more complex relations between them and their intermediate states, to finally obtain the output state of this second layer LSTM. All these LSTM layers use hyperbolic tangent functions to provide high non-linearities to the prediction model. Finally, the outputs of the second LSTM layer are processed by the dense output layer that is in charge of interrelating all these and performs a regression through which it predicts the point of interest. This dense output layer employs a linear activation, which is a widely used activation in regression tasks.

\begin{figure}[ht!]
\centering
\includegraphics[width=0.8\columnwidth]{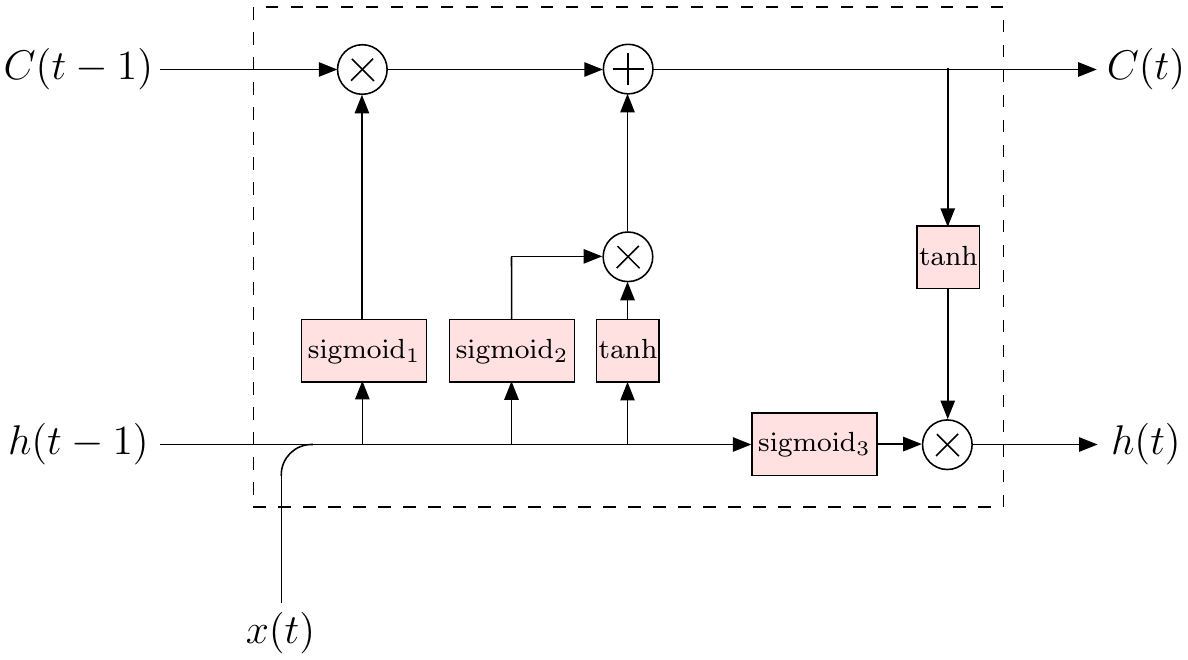}
\caption{Internal structure of LSTM blocks.}\label{fig:FLSTMblocks}
\end{figure}

The input time segment $\mathcal{I}$ with size $1,200$ (around 6 PPG signal cycles) is processed by the first LSTM layer that returns all the intermediate states to enrich the used information by the second LSTM layer; this second LSTM layer processes more complex and recurrent information to finally decide, along with the fully connected layer, what is the prediction of the point $\mathcal{P}(t_{c}+h_{p(n)})$, as we can see in Figure \ref{fig:FRNNdiag}. In Table \ref{tab:TDNNarch}, we can see the architecture in detail, with all its layers and parameters. Finally, to provide a better understanding of the used LSTM blocks inside, we can observe all its internal structure in Figure \ref{fig:FLSTMblocks}.

\subsubsection{Training}

\begin{figure*}[ht!]
\centering
\subfloat[]{\includegraphics[scale=0.46]{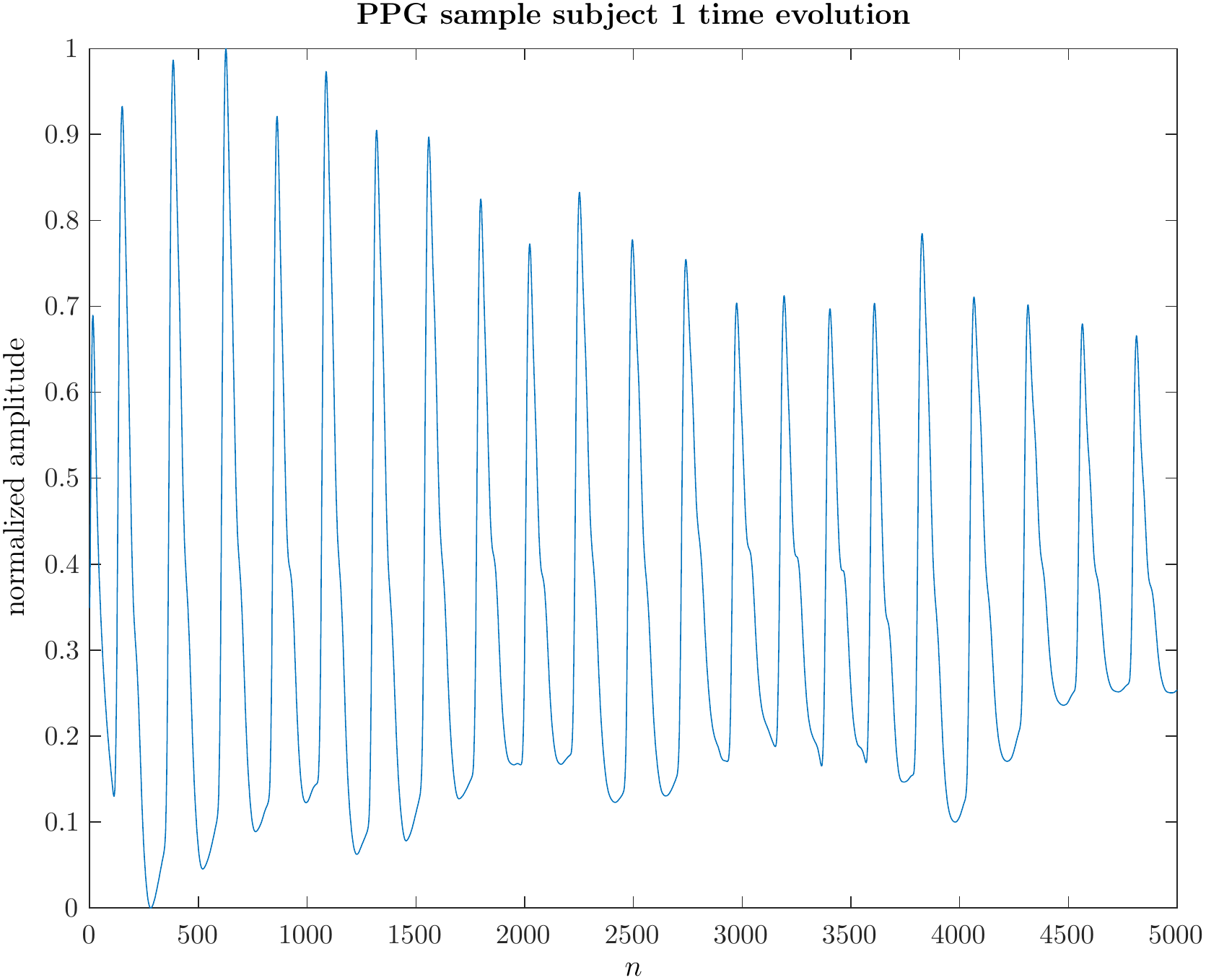}
\label{fig:FTEP1B}}
\subfloat[]{\includegraphics[scale=0.46]{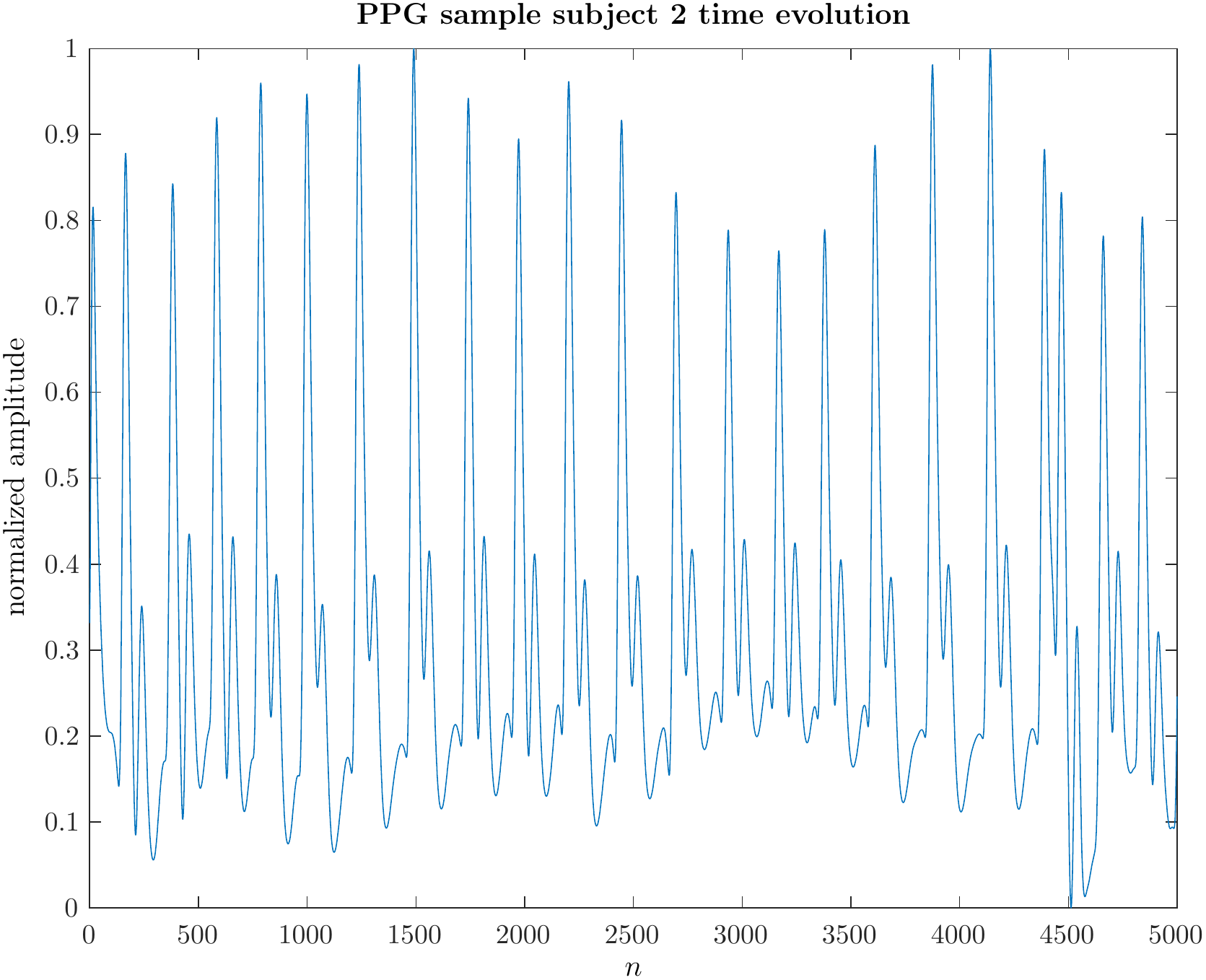}
\label{fig:FTEP2B}}\hfil
\subfloat[]{\includegraphics[scale=0.46]{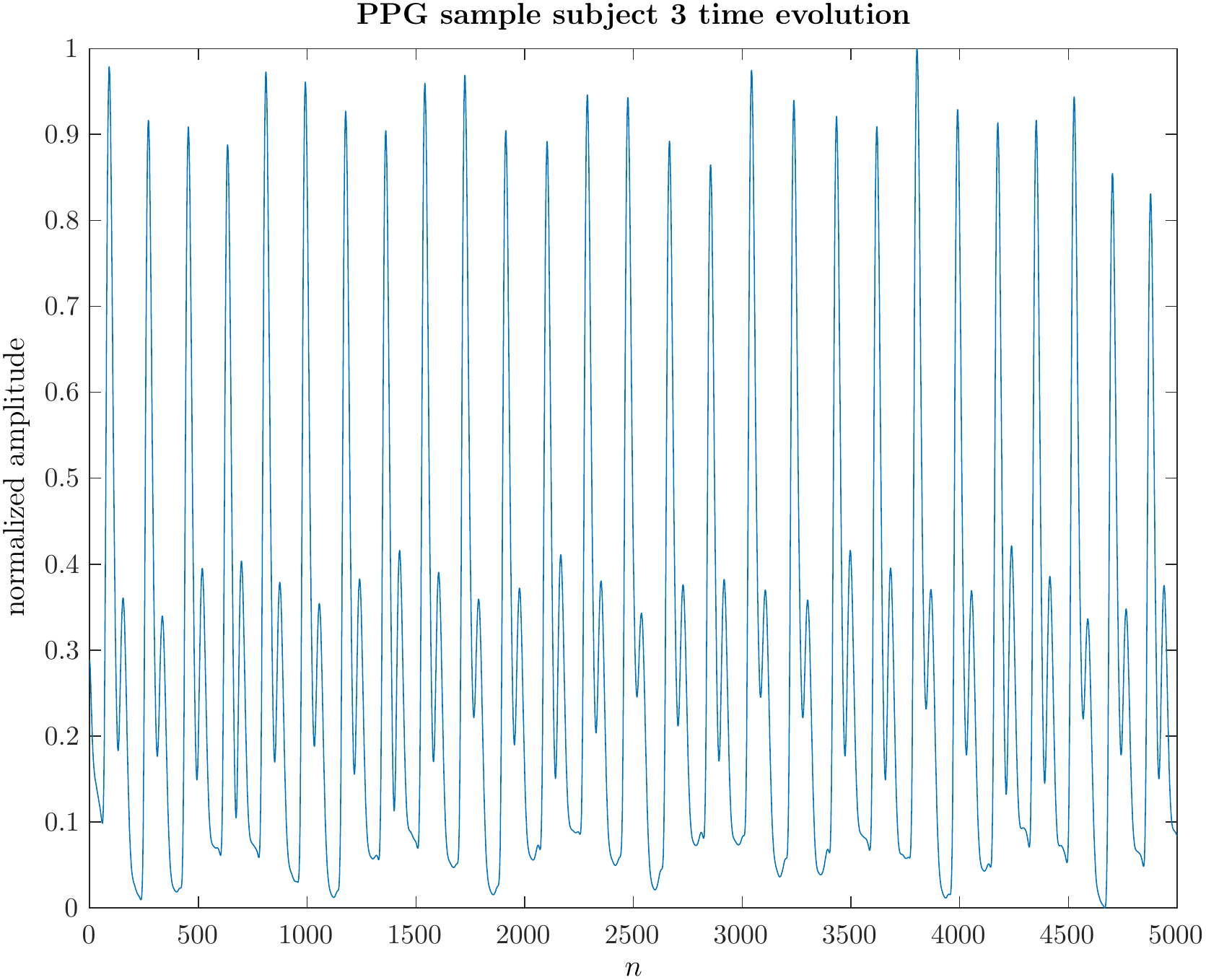}
\label{fig:FTEP3B}}
\subfloat[]{\includegraphics[scale=0.46]{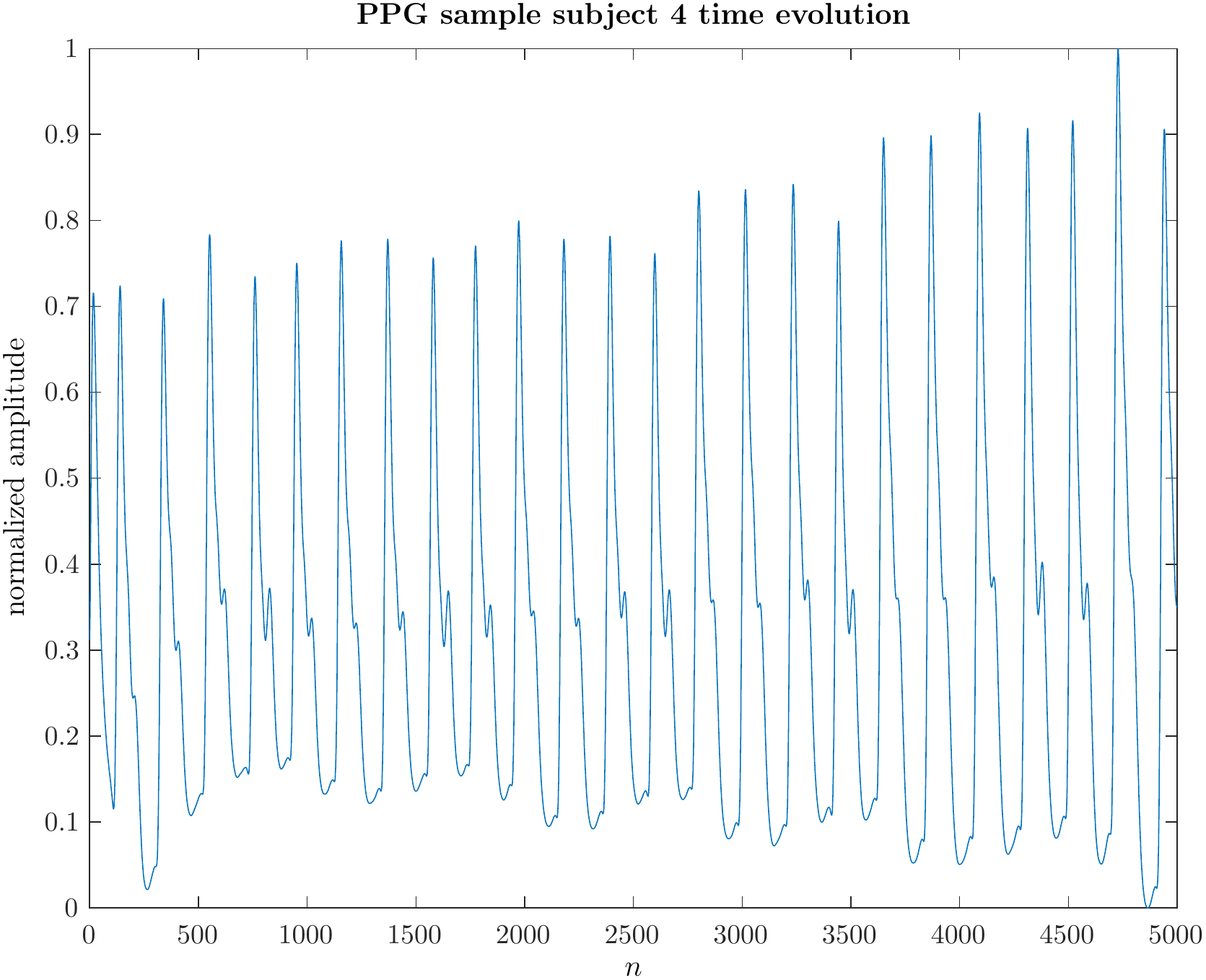}
\label{fig:FTEP4B}}\hfil
\subfloat[]{\includegraphics[scale=0.46]{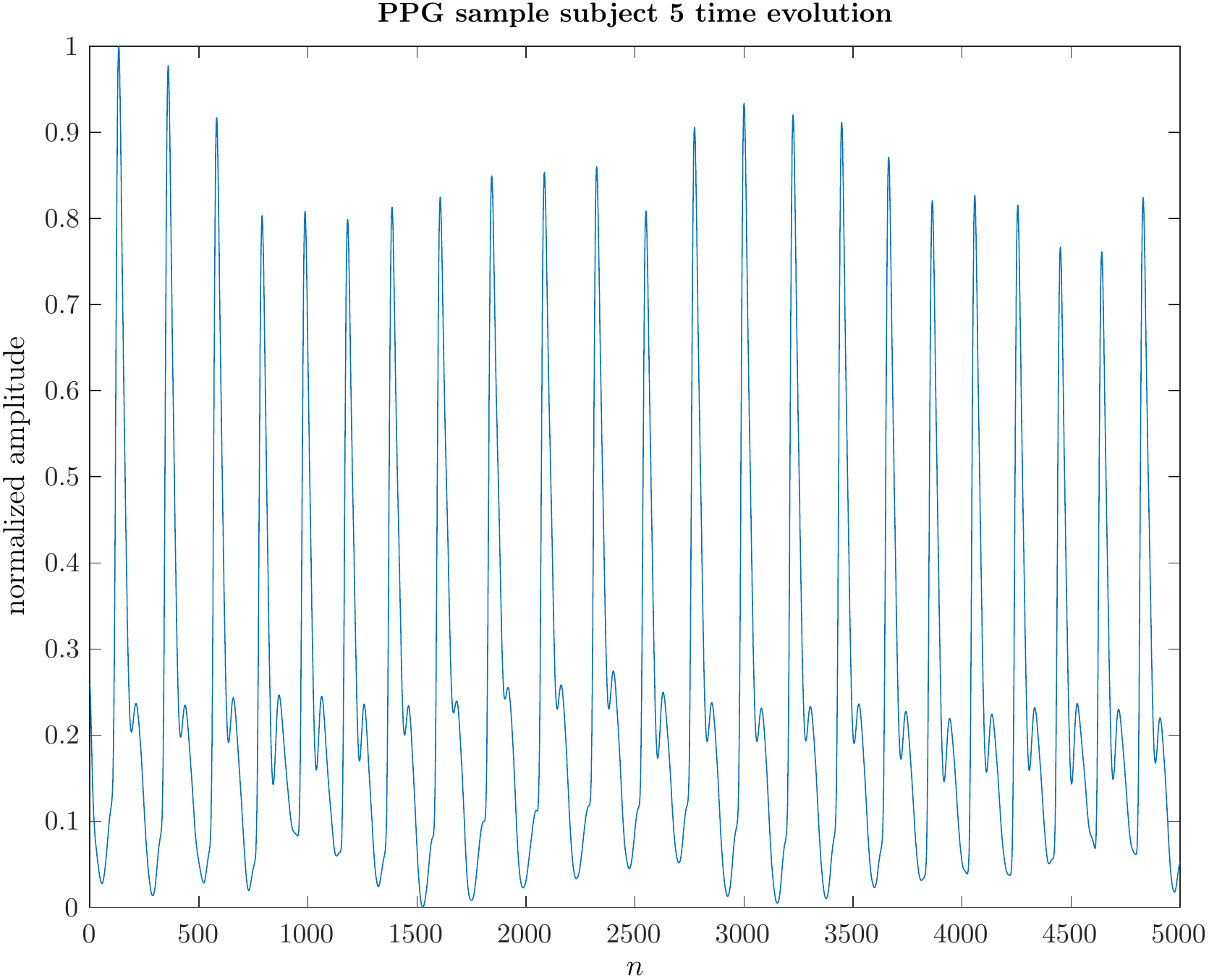}
\label{fig:FTEP5B}}
\caption{Real-world PPG signals (five individuals chosen at random): (\textbf{a}) PPG subject number 1 (PPG1); (\textbf{b}) PPG subject number 2 (PPG2); (\textbf{c}) PPG subject number 3 (PPG3); (\textbf{d}) PPG subject number 4 (PPG4); (\textbf{e}) PPG subject number 5 (PPG5).}
\label{fig:FPPGTE}
\end{figure*}

In the training step, we use the reference signals explained in \ref{sssec:syndata} to estimate the different trends horizon of prediction and characterize them. Besides that, we will train with five real-world PPG signals from the dataset, following \ref{sssec:realdata}, to provide an average PPG horizon of prediction. Each one of the processing of each one of the signals, will be composed of $N$ different trainings that will create $N$ different RNNs specialized in the prediction of each one of the defined points of interest to predict. So for each signal, we train $N$ RNNs. One hundred fifty thousand points form each one of the training signals. In Figures \ref{fig:FRSTE} and \ref{fig:FPPGTE}, we see the training signals conformed by the five most important dynamic behaviors and five real-world PPG signals selected randomly, respectively. In this step, we divide and train with the signals in a training set and a validation set composed by the 80\% and the 20\% per cent of the dataset, accordingly. These division ranges have been chosen since usually the training set constitutes 60 to 80 per cent of the dataset while the validation and test sets comprise 20 to 40 per cent. Let us know look in greater detail the training parameters.

\begin{enumerate}
\item \textbf{Optimizer}. The used optimizer is the same model applied to our CNN architecture, with the exception that the initial learning rate is $10^{-3}$ and, in this case, the used batch is 1000, because big batch sizes improve greatly the convergence, which is an already known problem in the recurrent neural networks structures. The total training time of the $N$ networks for each dynamical horizon is about 32 hours.

\item \textbf{Loss Function}. The RNN uses to train a training set formed by the input time segment and an output set conformed by the points to predict. The proposed regression loss function to use is the logarithm of the hyperbolic cosine or log-cosh loss. This loss tries to combine the advantages of the $L^{1}$-norm and $L^{2}$-norm, being robust to outliers at the same time that encourages the correct points to learn. The proposed log-cosh loss evaluates the differences between the predicted points $\mathcal{P}_{i}$ and the ground truth ones $\mathcal{Q}_{i}$, as we can see in Eq. \eqref{eqn:E02loss}.

\begin{equation}\label{eqn:E02loss}
L\left(\mathcal{Q},\mathcal{P}\right)=\sum_{i=1}^{n} \log \left(\cosh \left(\mathcal{Q}_{i}-\mathcal{P}_{i}\right)\right).
\end{equation}

\end{enumerate}

Just as we have done for our CNN architecture, we explain in depth and step by step how the training takes place for our RNN architecture. The first thing that needs to be commented is that it is necessary to train one RNN for each point of the prediction horizon to be predicted, that is to say, in order to find the prediction horizon at 50 points of view of a signal that we want to analyze, it is necessary to train 50 RNNs in charge of predicting each of the 50 points of interest. The fact of using a network for each point allows these networks to specialize in calculating the specific point. The inputs of the proposed RNN are composed of segments of 1200 points of the input signal, while the outputs of the same are composed of each of the 50 points following the input segment analyzed. The training of this network can be done both with real-world PPG signals and with previously generated synthetic signals. However, to build the prediction horizon solidly and to associate real PPG signals correctly to a trend, it is necessary to make the prediction horizon both from reference signals and from real-world PPG signals, thus being able to build a more robust and intuitive diagram. In the training process, random batches containing random segments of the signal to be analyzed is processed, with a length of 1,200 points. This process, randomize, provide a more generalized and learning-rich training. Once the batches are loaded, during 10 periods and through the use of the \textit{Adam} optimizer, the proposed system is trained to predict the output point correctly. For the training to be carried out, $80\%$ of each signal sample trained is used; $10\%$ of each signal for validation and $10\%$ of each signal for test. Once the prediction horizon and predicted points of the real-world PPG signals are obtained, the average prediction horizon measured by these PPG signals of interest is calculated.

\section{Results}\label{sec:Results}

Different experimental tests based on neural networks allow us for a qualitative and quantitative assessment of the dynamic behavior of a PPG signal at different timescales. The results are presented graphically and numerically. The numerical results shown in the tables concern all the measurements of the PPG signals available for each individual, i.e., around 600,000 points (40 minutes) per subject, while the figures only reflect time segments of a sample PPG signal.

\begin{figure}[ht!]
\centering
\includegraphics[width=0.94\columnwidth]{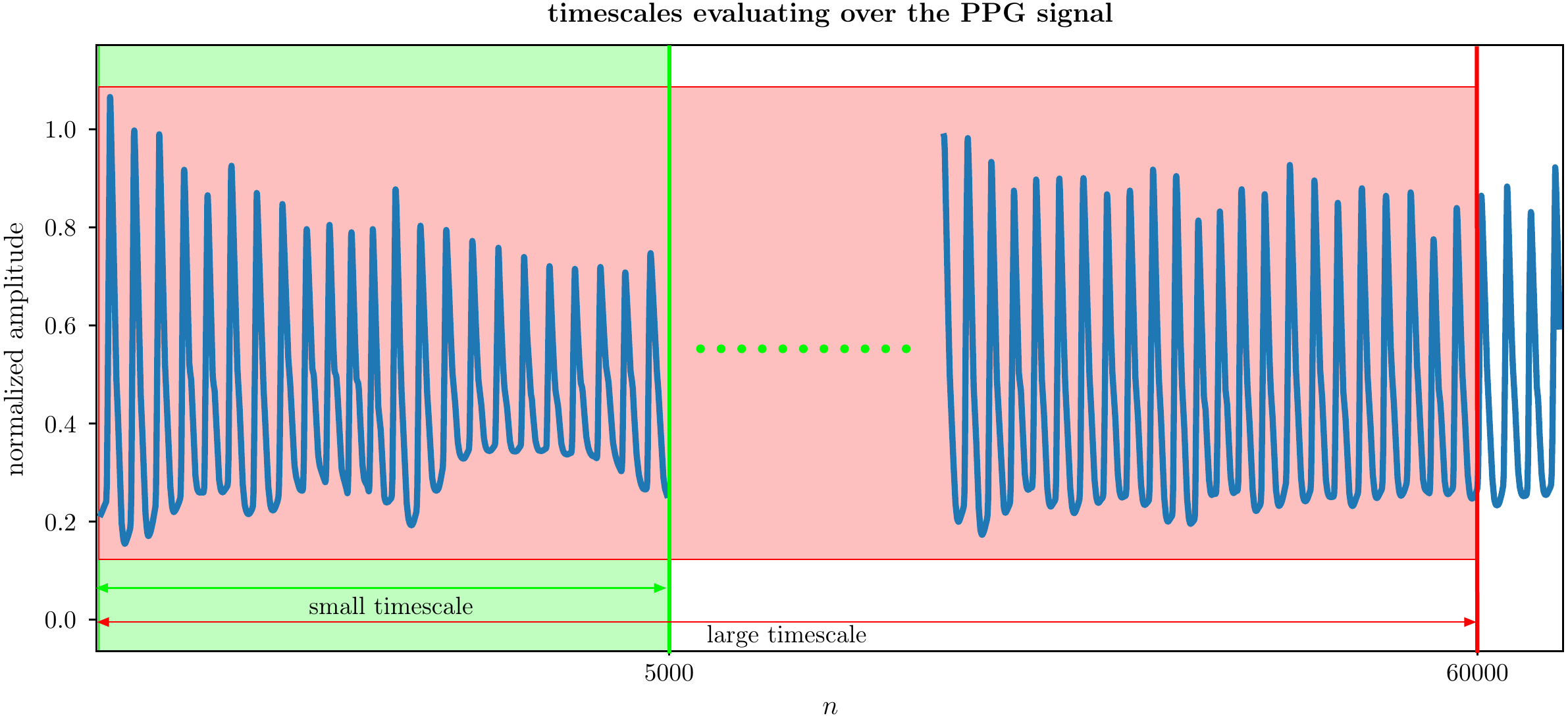}
\caption{Epochs of size 5,000 (small timescale) and 60,000 (large timescale) samples over a sample real-world PPG signal.}\label{fig:SLtrendsdiag}
\end{figure}  

\subsection{Dynamic behavior classification with a CNN model}\label{ssec:DBCCNN}

Here we show the evaluation of our implemented CNN-based dynamic behavior classification system (cf. \S\ \ref{ssec:CNNarch} for more details) using a real-world PPG signals dataset, consisting of 40 PPG signals from young and healthy individuals between the ages of 18 and 30, according to a national research project, and reference signals of well-known dynamics (cf. \S\ \ref{ssec:DATA}). Also, we look at the effect of noise and filtering on PPG signals to prove the classification capability of the system. The dynamic characterization process consists of two trained systems with a different number of input samples chosen empirically, the first with $\mathcal{I}=5,000$ and the second with $\mathcal{I}=60,000$. In this way, we provide two different perspectives: a very locally approach ($\mathcal{I}=5,000$),  or small timescale, and a less locally approach ($\mathcal{I}=60,000$), or a large timescale, of the system, as we can see in Figure \ref{fig:SLtrendsdiag}. The rationale for defining different timescales is that the dynamic behavior differs considerably on the PPG signal as we introduce more and more signal cycles, allowing us to discover the hidden dynamic richness of the PPG signal that we cannot appreciate at small timescales.

\subsubsection{Preliminary analysis of the PPG signal}\label{sssec:ourCNN}

\begin{figure}[ht!]
\centering
\subfloat[]{\includegraphics[width=0.48\columnwidth]{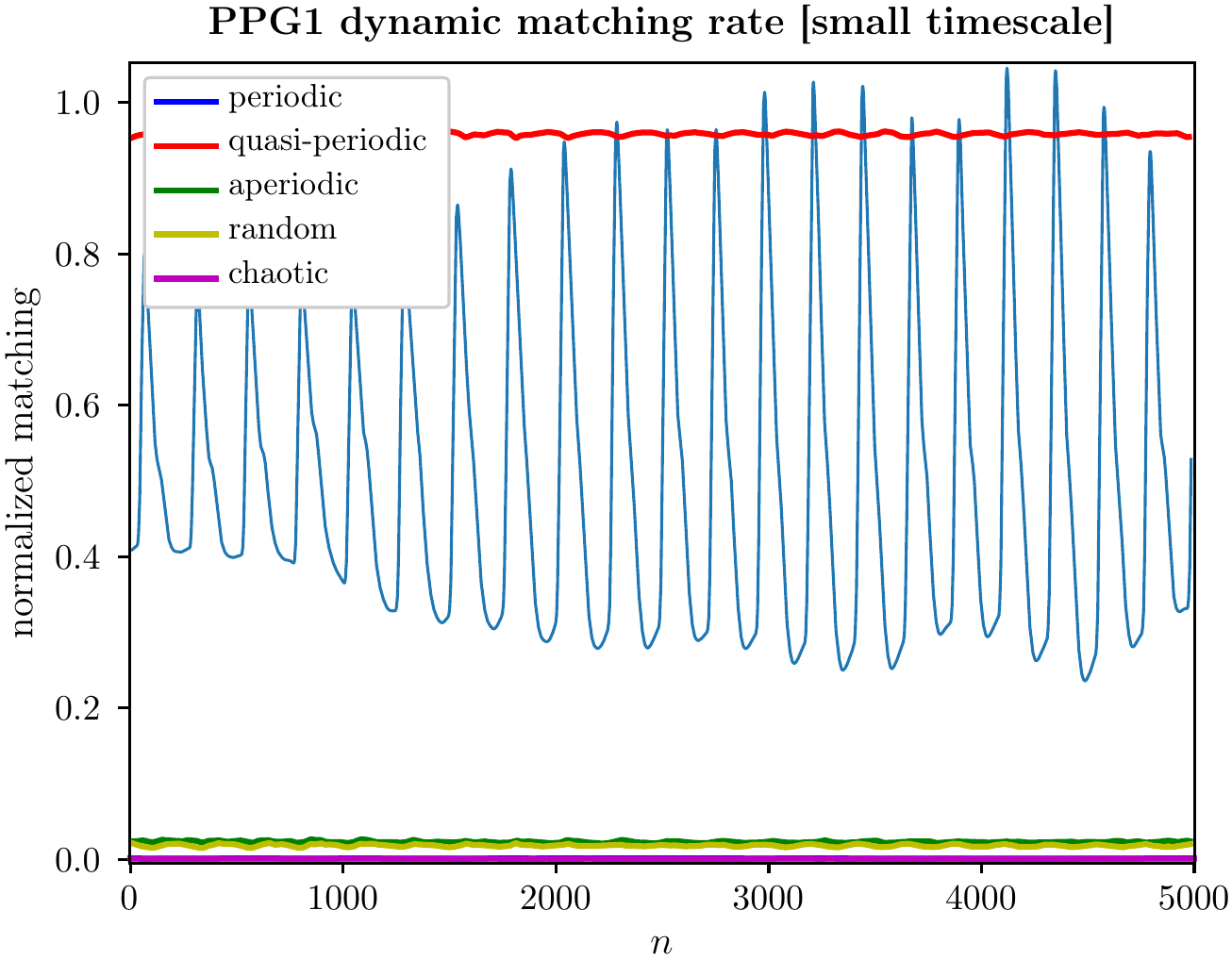}
\label{fig:FCNNP1BST}}
\subfloat[]{\includegraphics[width=0.48\columnwidth]{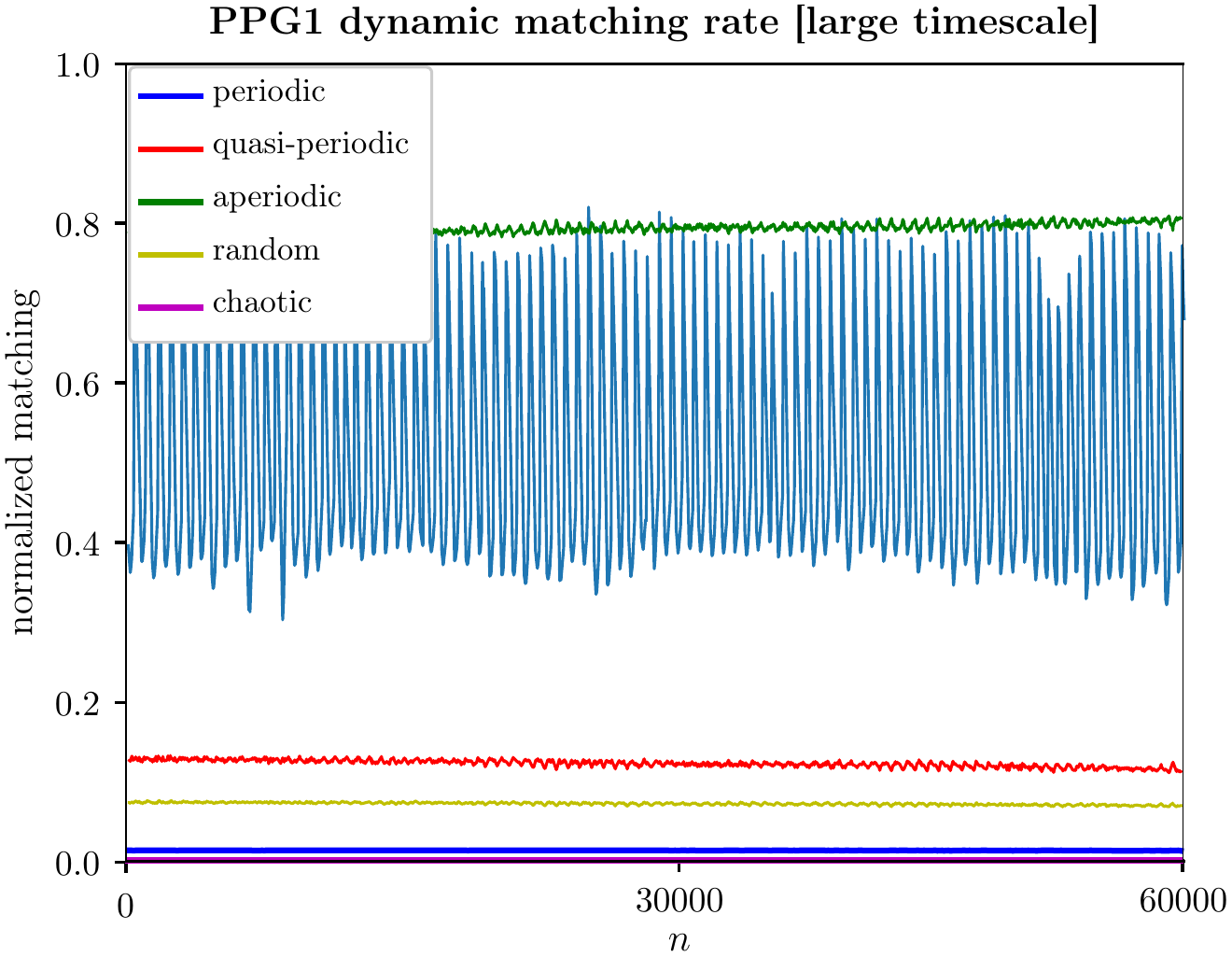}
\label{fig:FCNNP1BLT}}
\caption{From a sample PPG signal (subject number 1): (\textbf{a}) Dynamic behavior on a small timescale with our proposed CNN architecture; (\textbf{b}) Dynamic behavior on a large timescale with identical architecture.}
\label{fig:FCNNP1B}
\end{figure}

This first experimental test allows us to analyze the dynamic behavior of PPG signals at different time scales. Unless stated otherwise, to avoid high-frequency noise and to some extent, motion artifacts, all PPG signals are filtered with a simple Butterworth bandpass filter with cutoff frequencies at 0.01 and 8 Hz. The notion of small timescale means using as input to our CNN model $\mathcal{I}=5,000$ samples, and large timescale means $\mathcal{I}=60,000$ samples. In Figure \ref{fig:FCNNP1B}, we show an  input sample PPG signal superposed with the dynamic behaviors matching-up rate allocated by our proposed CNN model at different timescales. In Table \ref{tab:CNNresults}, we refer to the corresponding numerical values, also providing the average values of each dynamic behavior for all PPG signals.

As can be seen in Table \ref{tab:CNNresults}, at small timescales, the predominant dynamic behaviour of PPG signals is quasi-periodic, with matching-up rate of around 99\%, and residual rates for the rest of the dynamics. However, at larger timescales, the dynamics get more complex and intriguing. It is now the aperiodic evolution that manages the dynamic behavior of the PPG signal, with a by no means negligible quasi-periodic component and a weakest random component.

\setcellgapes{5pt}\makegapedcells
\begin{table}[ht!]
\caption{Dynamic composition of PPG signals, expressed as a matching-up percentage of reference signals.}\label{tab:CNNresults}
\centering
\begin{adjustbox}{max width=\linewidth}
\begin{tabular}{|c|L|L|L|L|L|}
\hline
\multicolumn{6}{|c|}{\textsc{small timescales}} \\
\hline
\textbf{Signal} &  \multicolumn{1}{c|}{\textbf{Periodic (\%)}} & \multicolumn{1}{c|}{\textbf{Quasi-periodic (\%)}} & \multicolumn{1}{c|}{\textbf{Aperiodic (\%)}} & \multicolumn{1}{c|}{\textbf{Random (\%)}} & \multicolumn{1}{c|}{\textbf{Chaotic (\%)}} \\	
\hline
PPG1 & 0.01 & 99.88 & 0.05 & 0.03 & 0.03 \\
\hline
PPG2 & 0.01 & 99.89 & 0.04 & 0.02 & 0.03 \\
\hline
PPG3 & 0.01  & 99.87 & 0.04 & 0.03 &  0.03 \\
\hline
PPG4 &  0.01 & 99.88 & 0.05 & 0.02 & 0.04  \\
\hline
Average of all (40) PPG signals &  0.01 & 99.88 & 0.04 & 0.02 &  0.03 \\
\hline
\multicolumn{6}{|c|}{\textsc{large timescales}} \\
\hline
\textbf{Signal} &  \multicolumn{1}{c|}{\textbf{Periodic (\%)}} & \multicolumn{1}{c|}{\textbf{Quasi-periodic (\%)}} & \multicolumn{1}{c|}{\textbf{Aperiodic (\%)}} & \multicolumn{1}{c|}{\textbf{Random (\%)}} & \multicolumn{1}{c|}{\textbf{Chaotic (\%)}} \\	
\hline
PPG1 & 1.26 & 11.43 & 80.51 & 6.78 & 0.02 \\
\hline
PPG2 &  1.24 & 11.04 & 81.05 & 6.64 & 0.02 \\
\hline
PPG3 & 1.23  & 10.51 & 81.79 & 6.45 &  0.02 \\
\hline
PPG4 & 1.28  & 23.11 & 65.47 & 10.11 & 0.02  \\
\hline
Average of all (40) PPG signals & 1.26  & 14.02 & 77.20 & 7.50 & 0.02  \\
\hline
\end{tabular}
\end{adjustbox}
\end{table}

\setcellgapes{5pt}\makegapedcells
\begin{table}[ht!]
\caption{Average value of the dynamic composition of all (40) PPG signals, expressed as a matching-up percentage of reference signals, for raw and filtered PPG signals.}\label{tab:CNNRFresults}
\centering
\begin{adjustbox}{max width=\linewidth}
\begin{tabular}{|c|L|L|L|L|L|}
\hline
\multicolumn{6}{|c|}{\textsc{small timescales}} \\
\hline
\textbf{Signal} &  \multicolumn{1}{c|}{\textbf{Periodic (\%)}} & \multicolumn{1}{c|}{\textbf{Quasi-periodic (\%)}} & \multicolumn{1}{c|}{\textbf{Aperiodic (\%)}} & \multicolumn{1}{c|}{\textbf{Random (\%)}} & \multicolumn{1}{c|}{\textbf{Chaotic (\%)}} \\	
\hline
Raw PPG signals & 0.36 & 76.79 & 18.76 & 3.75 & 0.33 \\
\hline
Filtered PPG signals & 0.13 & 96.02 & 2.11 & 1.65 & 0.09 \\
\hline
$|\Delta|$ & 0.23 & 19.23 & 16.65 & 2.1 & 0.24 \\
\hline
\multicolumn{6}{|c|}{\textsc{large timescales}} \\
\hline
\textbf{Signal} &  \multicolumn{1}{c|}{\textbf{Periodic (\%)}} & \multicolumn{1}{c|}{\textbf{Quasi-periodic (\%)}} & \multicolumn{1}{c|}{\textbf{Aperiodic (\%)}} & \multicolumn{1}{c|}{\textbf{Random (\%)}} & \multicolumn{1}{c|}{\textbf{Chaotic (\%)}} \\	
\hline
Raw PPG signals & 0.35 & 75.71 & 17.20 & 6.73 & 0.00  \\
\hline
Filtered PPG signals & 0.95  & 3.21 & 92.74 & 3.07 & 0.01 \\
\hline
$|\Delta|$ &  0.60 & 72.5 & 75.54 & 3.66 & 0.01 \\
\hline
\end{tabular}
\end{adjustbox}
\end{table}

\subsubsection{Raw versus filtered PPG signals}\label{sssec:filter}

\begin{figure}[ht!]
\centering
\subfloat[]{\includegraphics[width=0.475\columnwidth]{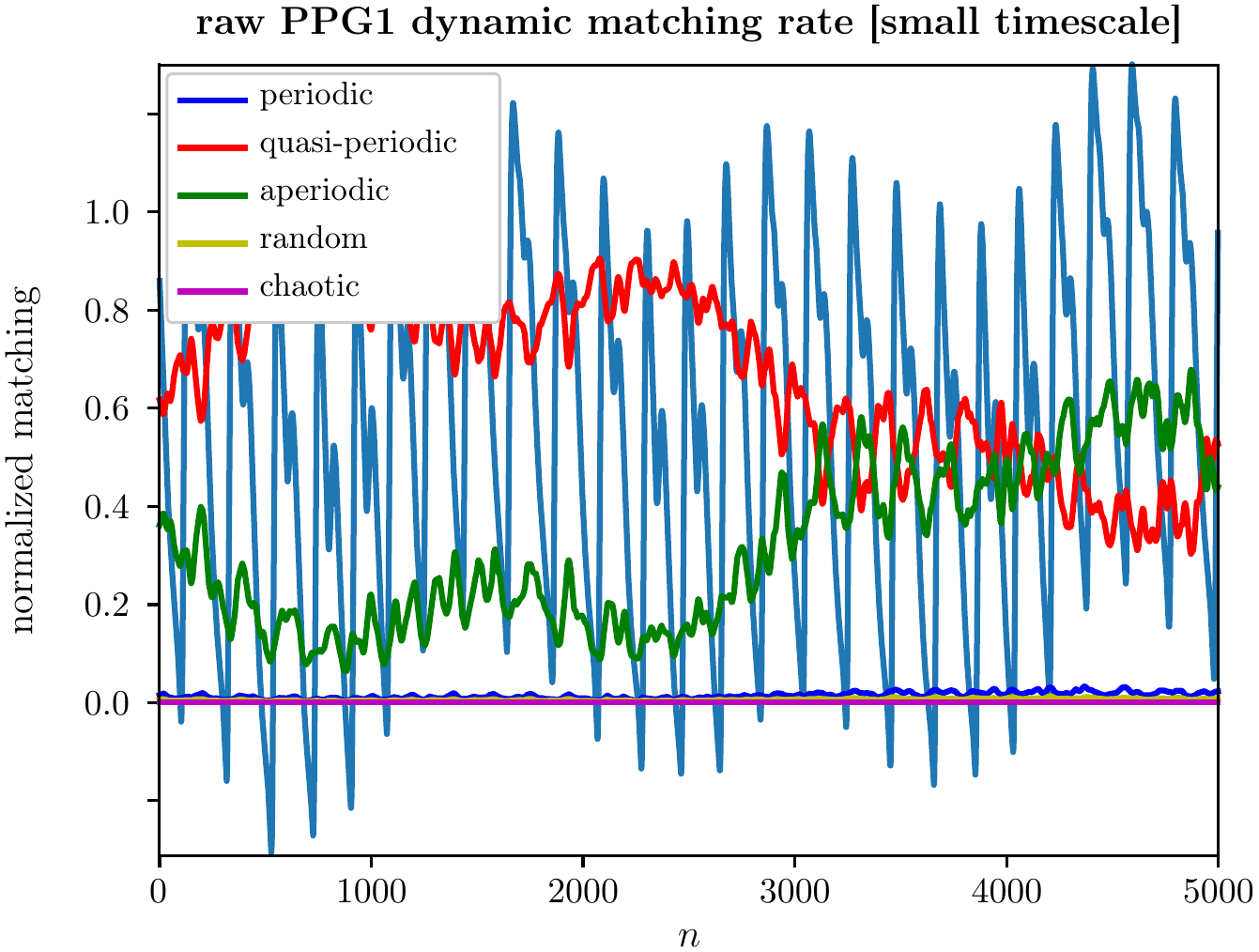}
\label{fig:FCNNRST}}
\subfloat[]{\includegraphics[width=0.475\columnwidth]{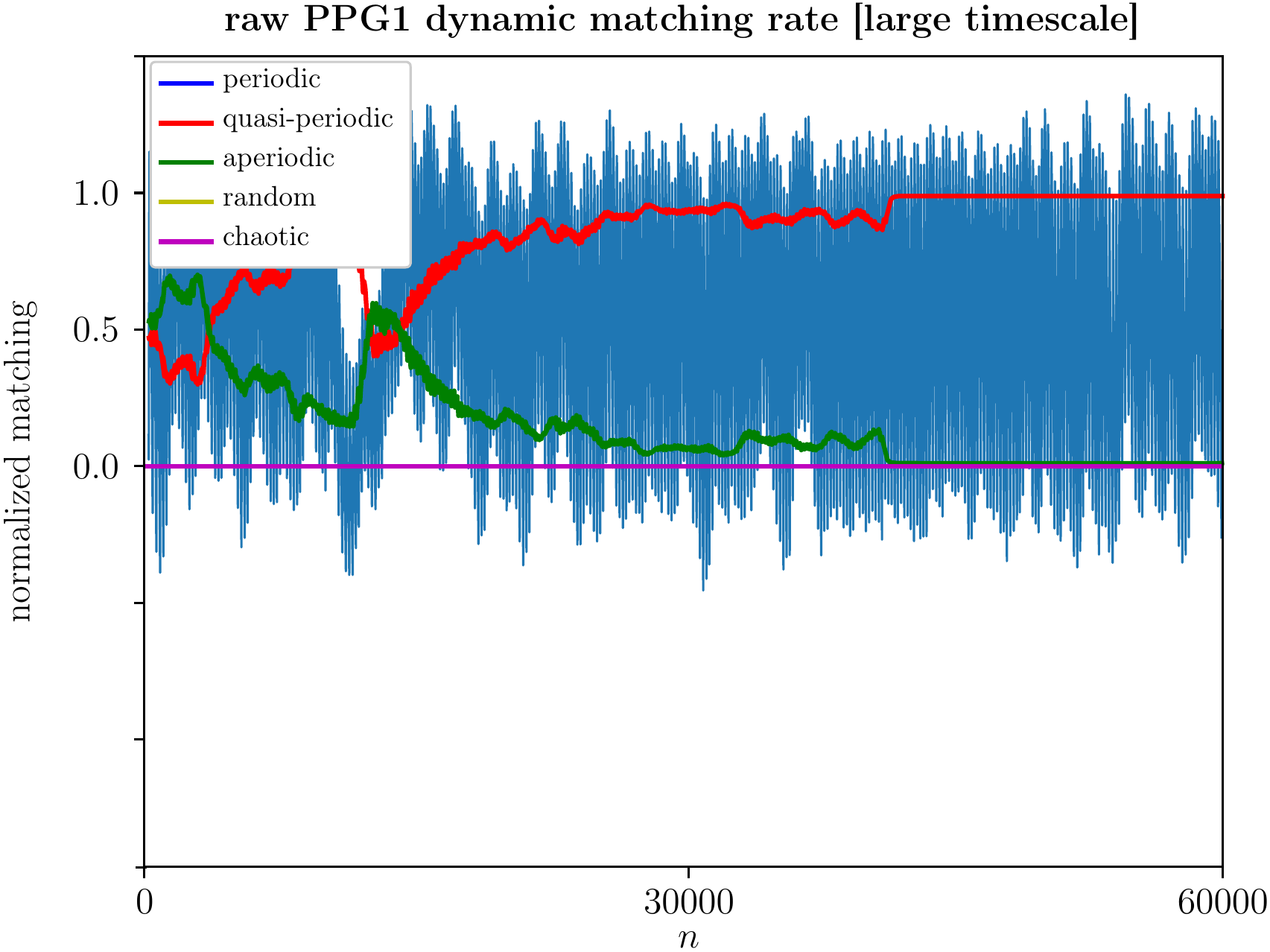}
\label{fig:FCNNRLT}}\hfil
\subfloat[]{\includegraphics[width=0.47\columnwidth]{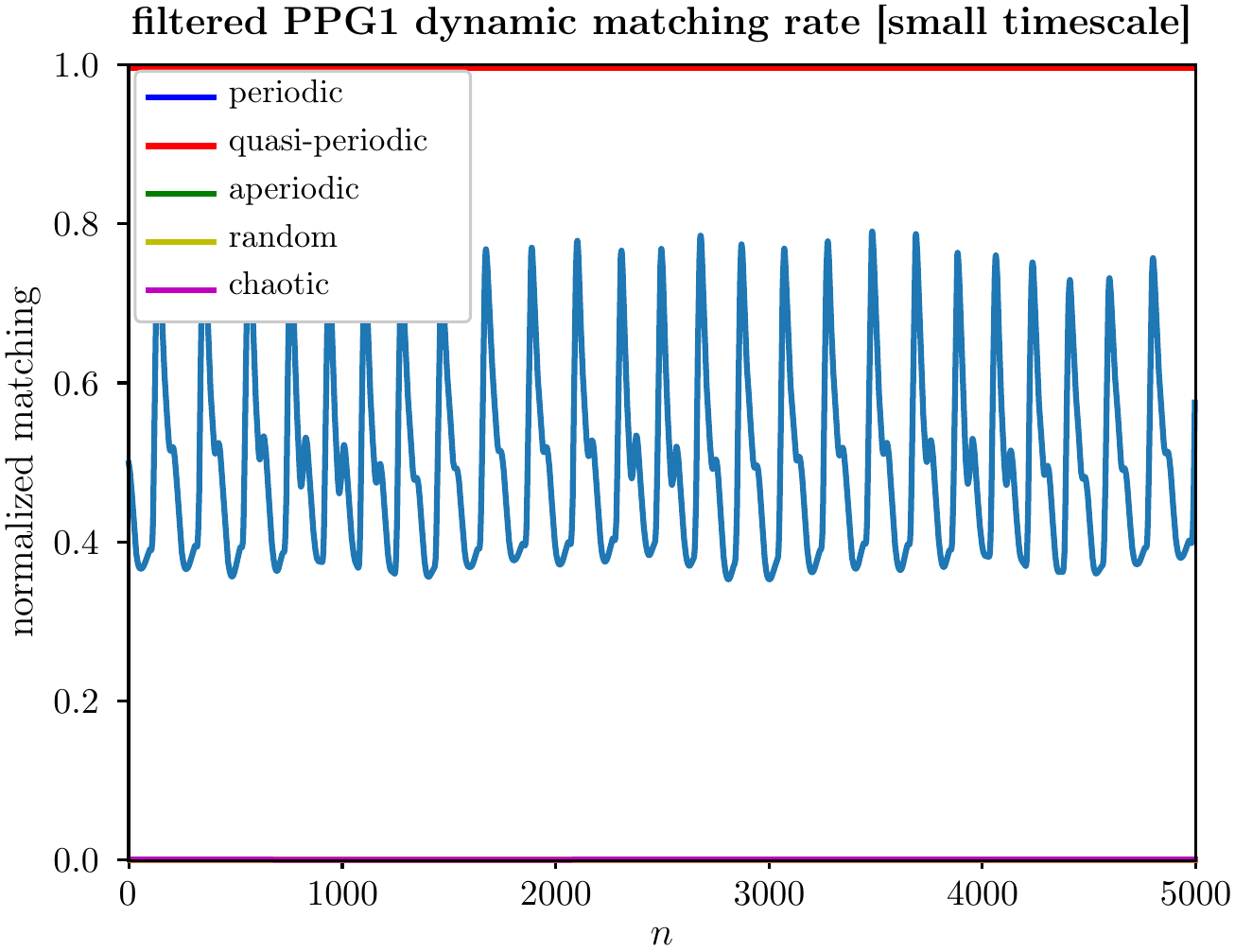}
\label{fig:FCNNFST}}
\subfloat[]{\includegraphics[width=0.495\columnwidth]{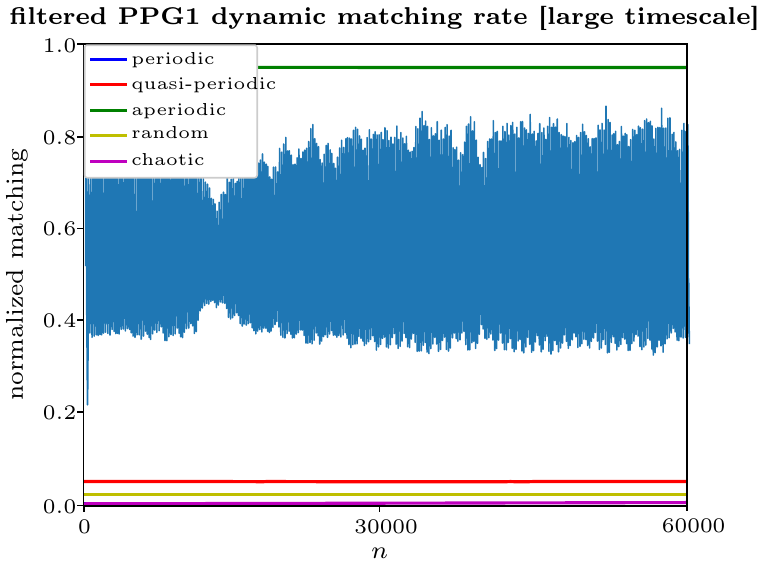}
\label{fig:FCNNFLT}}
\caption{From a sample PPG signal (subject number 1): (\textbf{a}) Dynamic behavior on a small timescale with our proposed CNN architecture for the raw PPG signal; (\textbf{b}) Dynamic behavior on a large timescale with identical architecture for the raw PPG signal; (\textbf{c}) Dynamic behavior on a small timescale with our proposed CNN architecture for the filtered PPG signal; (\textbf{d}) Dynamic behavior on a large timescale with identical architecture for the filtered PPG signal.}
\label{fig:FCNNRF}
\end{figure}

The next experimental test aims us at assessing the discriminant power of the dynamic behavior classifier implemented by our CNN model for filtered and raw (unfiltered) PPG signals. Raw PPG signals include those supplied direct by the research project itself. Although we are aware that PPG signals have some preprocessing, it is our understanding that PPG signals have quite a few motion artifacts, among other limiting factors. For filtered PPG signals, we simply kept that stated in \ref{sssec:ourCNN}, according to which all PPG signals are filtered with a Butterworth bandpass filter with cutoff frequencies at 0.01 and 8 Hz, in order to remove artifacts while retaining as far as possible all dynamic information available. As in the previous experimental test, we explore the dynamic differences between raw and filtered PPG signals at different timescales. In Figure \ref{fig:FCNNRF}, we show an input sample raw and filtered PPG signal superposed with the dynamic behaviors matching-up rate allocated by our proposed CNN model at different timescales. In Table \ref{tab:CNNRFresults}, we refer to the corresponding numerical values, also providing the average values of each dynamic behavior for all PPG signals, in conjunction with the percentage change $|\Delta|$, in absolute terms, between filtered and raw signals.

In accordance with Table \ref{tab:CNNRFresults}, we can see how filtering enhances the discriminating power of our CNN architecture. Raw PPG signals enter much noise into the outputs provided by our CNN model, especially on a large timescale, where even the predominant dynamics, quasi-periodic and aperiodic behaviors, exchange their role with respect to what happens with filtered signals. At small timescales, the effect is not as marked, perhaps due to possible noise in the input signal or spurious artifacts, and the outputs provided by our CNN model are more stable. However, the percentage change is still significant in the case of the predominant dynamic behaviors, more than 15\%, and the classification process implemented may make mistakes. 

\subsubsection{PPG signals with added white noise}\label{sssec:noise}

\begin{figure}[ht!]
\centering
\subfloat[]{\includegraphics[width=0.3325\columnwidth]{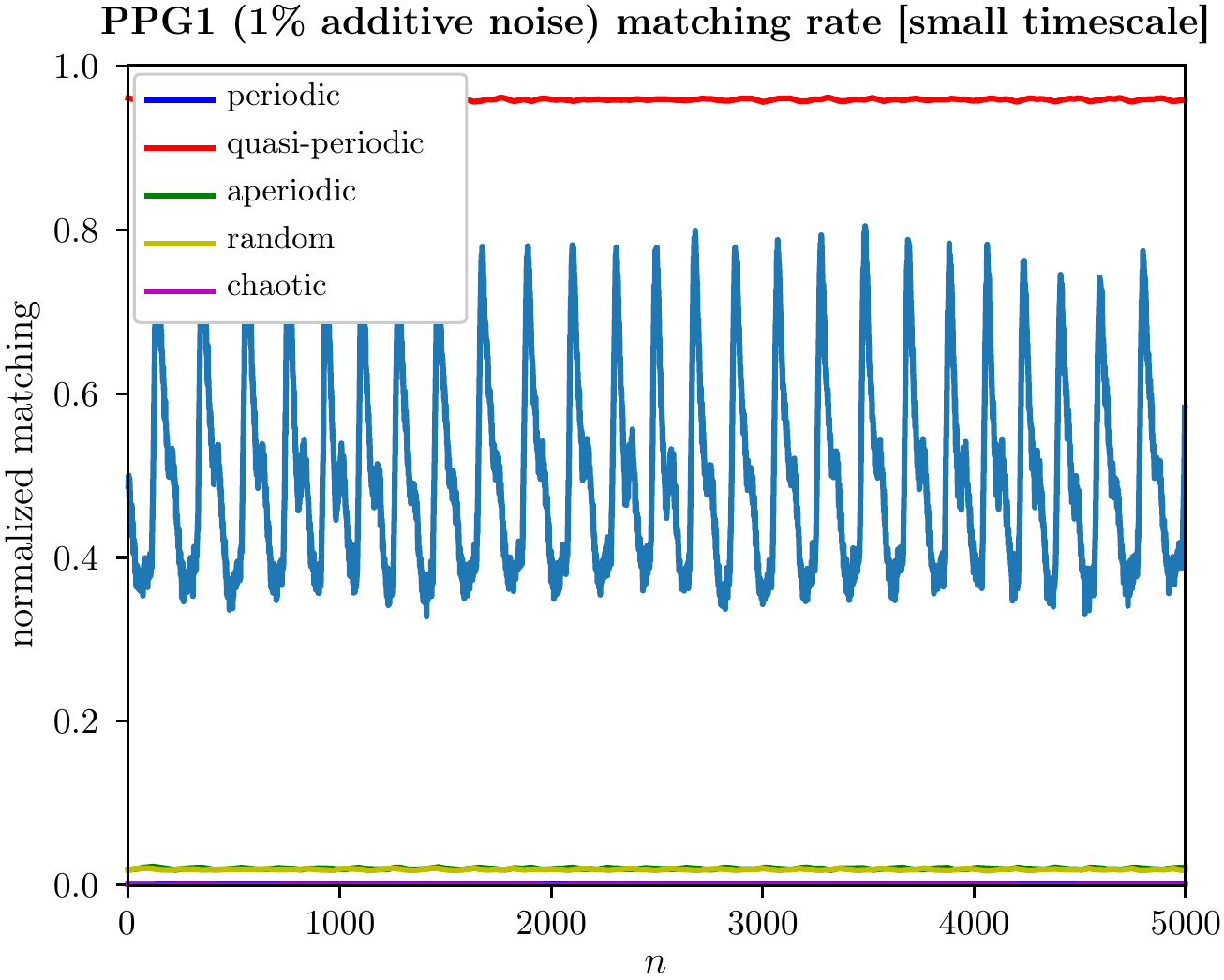}
\label{fig:FCNNWNSTA1}}
\subfloat[]{\includegraphics[width=0.33\columnwidth]{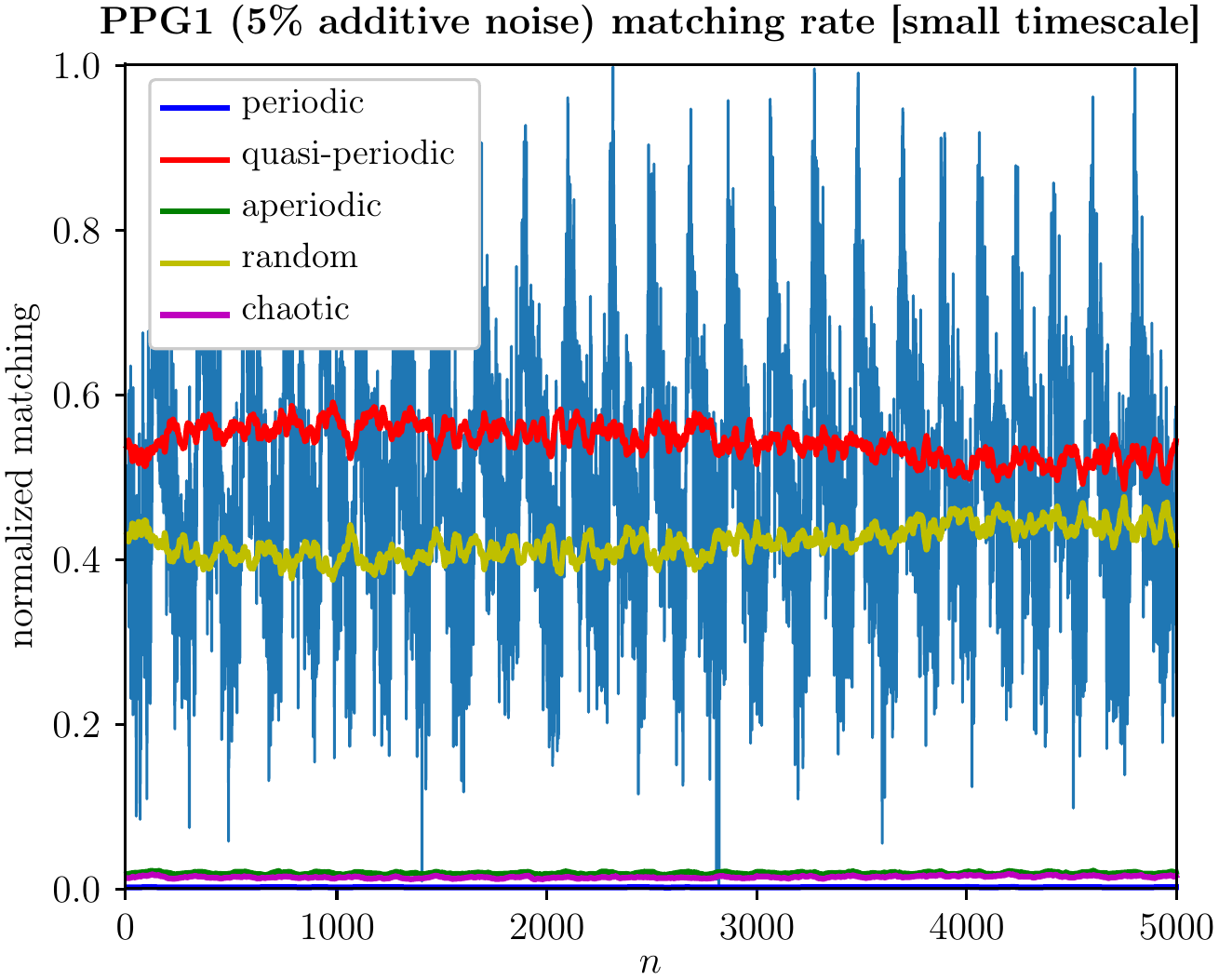}
\label{fig:FCNNWNSTA2}}
\subfloat[]{\includegraphics[width=0.33\columnwidth]{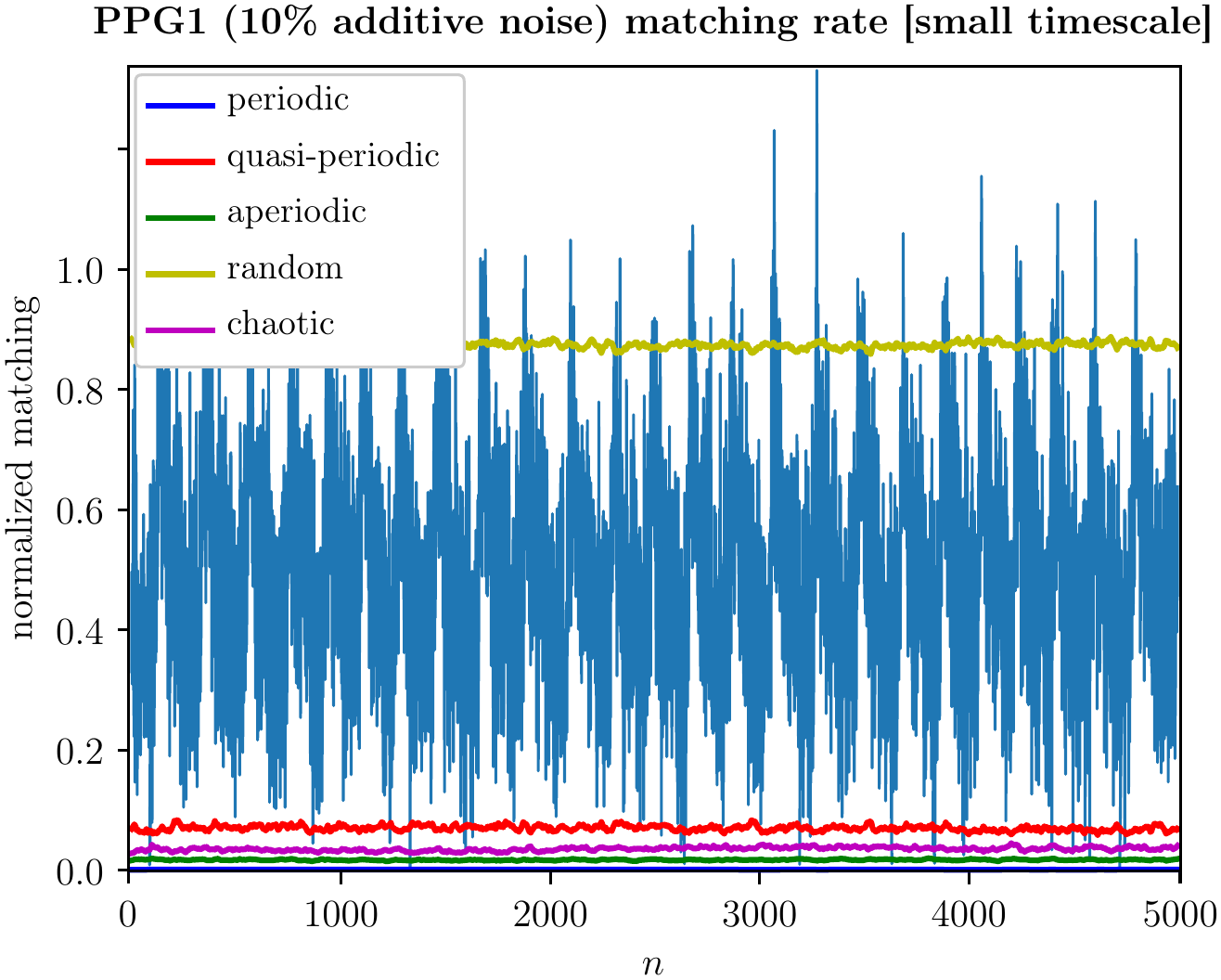}
\label{fig:FCNNWNSTA3}}\hfil
\subfloat[]{\includegraphics[width=0.33\columnwidth]{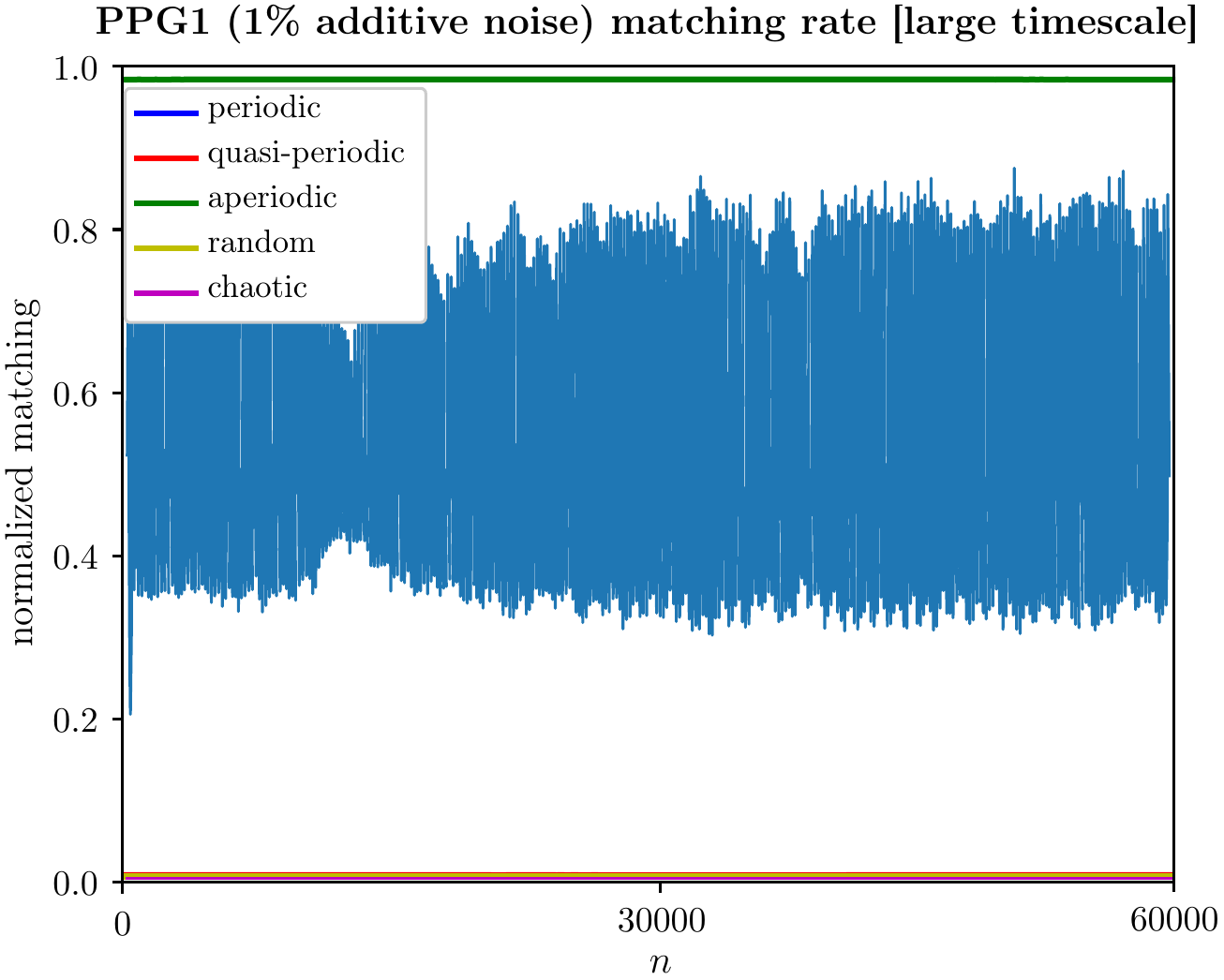}
\label{fig:FCNNWNLTA1}}
\subfloat[]{\includegraphics[width=0.3375\columnwidth]{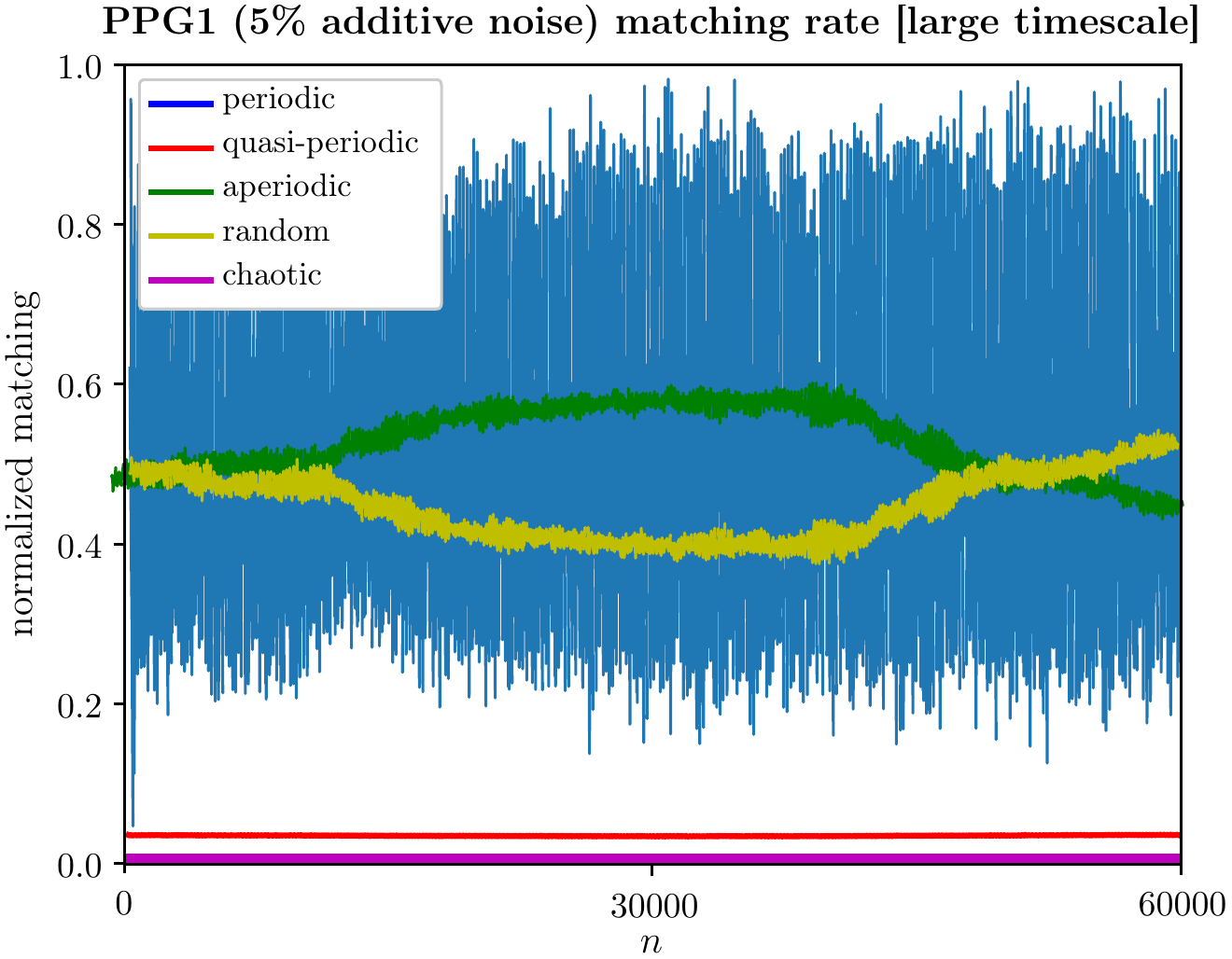}
\label{fig:FCNNWNLTA2}}
\subfloat[]{\includegraphics[width=0.335\columnwidth]{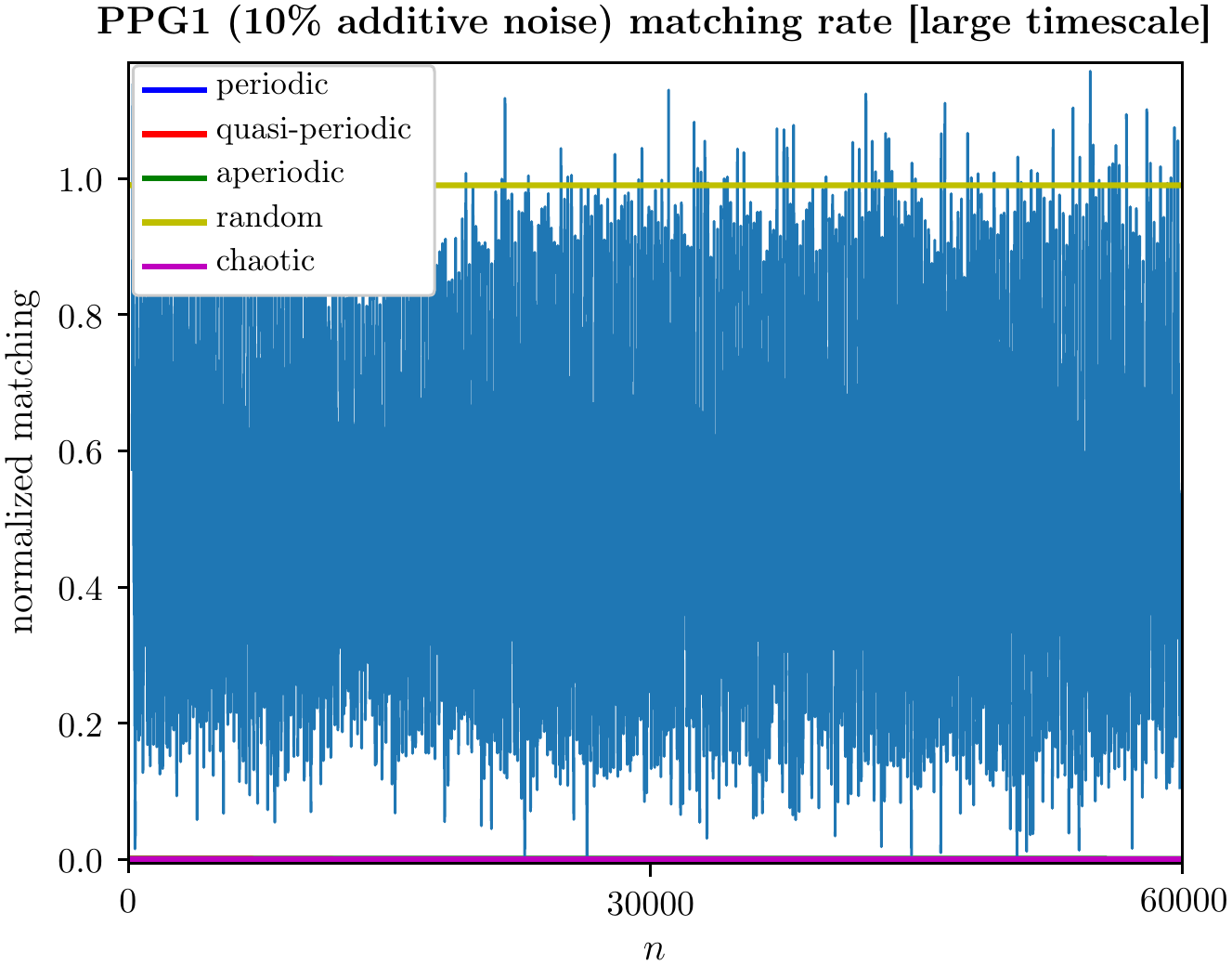}
\label{fig:FCNNWNLTA3}}
\caption{From a sample PPG signal (subject number 1): (\textbf{a})-(\textbf{c}) Dynamic behavior on a small timescale with our proposed CNN architecture for the PPG signal at 1\%, 5\% and 10\% noise level; (\textbf{d})-(\textbf{f}) Dynamic behavior on a large timescale with oure proposed CNN architecture for the PPG signal at 1\%, 5\% and 10\% noise level.}
\label{fig:FCNNWN}
\end{figure}

Within our CNN architecture, the latest experimental test probes the robustness of the dynamic behavior classifier when examined PPG signals are noisy. That is why we add different additive white noise levels to real-world PPG signals, and roughly estimate the tolerance threshold of our CNN network to discriminate dynamic behaviors in noisy PPG signals at different timescales. In Figure \ref{fig:FCNNWN}, we show an input sample noisy PPG signal, at 1\%, 5\% and 10\% noise level, superposed with the dynamic behaviors matching-up rate allocated by our proposed CNN model at different timescales. In Table \ref{tab:CNNWNresults}, we refer to the corresponding numerical values, also providing the average values of each dynamic behavior for all PPG signals, according to the added white noise amplitude.

\setcellgapes{5pt}\makegapedcells
\begin{table}[ht!]
\caption{Dynamic composition of all (40) PPG signals, expressed as a matching-up percentage of reference signals, for different noise levels.}\label{tab:CNNWNresults}
\centering
\begin{adjustbox}{max width=\linewidth}
\begin{tabular}{|L|L|L|L|L|L|}
\hline
\multicolumn{6}{|c|}{\textsc{small timescales}} \\
\hline
\multicolumn{1}{|c|}{\textbf{Additive noise maximal amplitude (\%)}} &  \multicolumn{1}{c|}{\textbf{Periodic (\%)}} & \multicolumn{1}{c|}{\textbf{Quasi-periodic (\%)}} & \multicolumn{1}{c|}{\textbf{Aperiodic (\%)}} & \multicolumn{1}{c|}{\textbf{Random (\%)}} & \multicolumn{1}{c|}{\textbf{Chaotic (\%)}} \\	
\hline
1 & 0.13 & 95.91 & 1.99 & 1.87 & 0.10 \\
\hline
5 & 0.19 & 54.37 & 1.94 & 42.03 & 1.46 \\
\hline
10 & 0.04 & 3.35 & 1.03 & 92.14 & 3.44 \\
\hline
\multicolumn{6}{|c|}{\textsc{large timescales}} \\
\hline
\multicolumn{1}{|c|}{\textbf{Additive noise maximal amplitude (\%)}} &  \multicolumn{1}{|c|}{\textbf{Periodic (\%)}} & \multicolumn{1}{c|}{\textbf{Quasi-periodic (\%)}} & \multicolumn{1}{c|}{\textbf{Aperiodic (\%)}} & \multicolumn{1}{c|}{\textbf{Random (\%)}} & \multicolumn{1}{c|}{\textbf{Chaotic (\%)}} \\	
\hline
1 & 0.00 & 0.19 & 99.77 & 0.02 & 4.50 \\
\hline
5 & 0.19  & 2.88 & 52.88 & 44.03 & 0.00 \\
\hline
10 & 0.01  & 0.06 & 0.09 & 99.83 &7.69  \\
\hline
\end{tabular}
\end{adjustbox}
\end{table}

Within Table \ref{tab:CNNWNresults}, we can see that with 5\% white noise added to PPG signals, our CNN architecture already detects a strong random component at the expense of the alleged winning dynamic choice in each timescale. An interesting point is that with 1\% added white noise, the random component is hidden, and at the same time the chaotic component resurfaces, but not above 4.5\% on a large timescale.

\begin{figure}[ht!]
\centering
\subfloat[]{\includegraphics[width=0.3325\columnwidth]{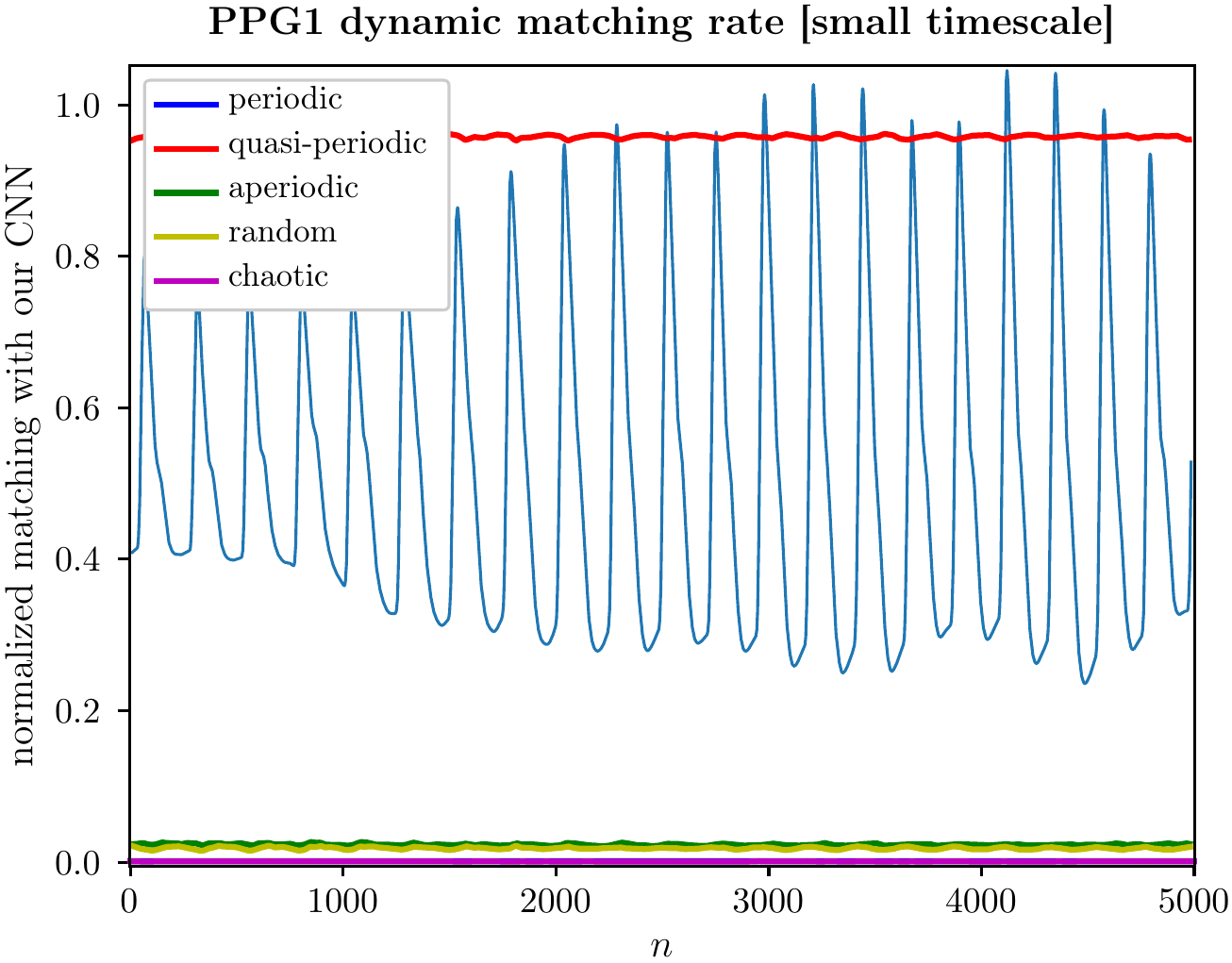}
\label{fig:FCNNDASTA1}}
\subfloat[]{\includegraphics[width=0.335\columnwidth]{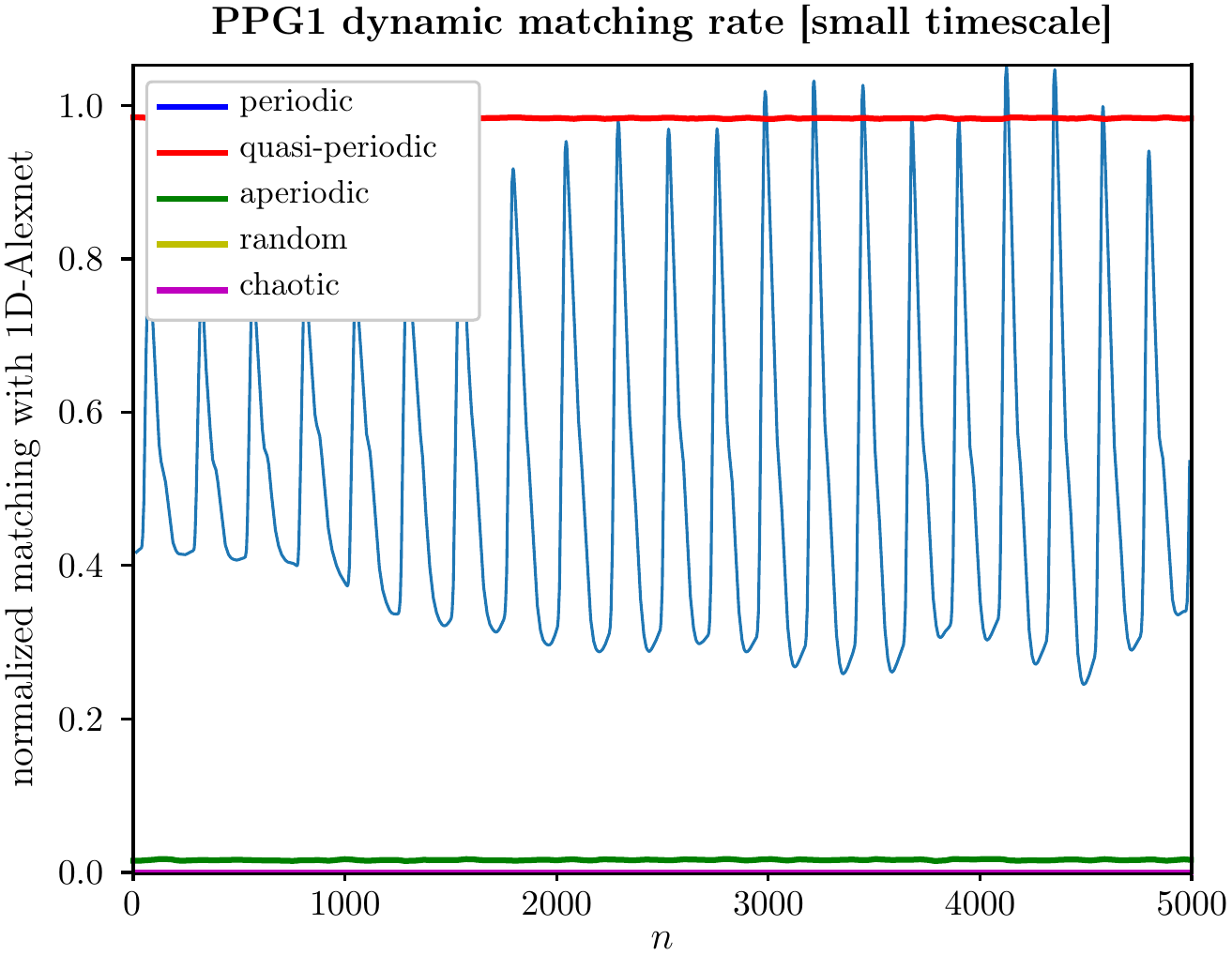}
\label{fig:FCNNDASTA2}}
\subfloat[]{\includegraphics[width=0.33\columnwidth]{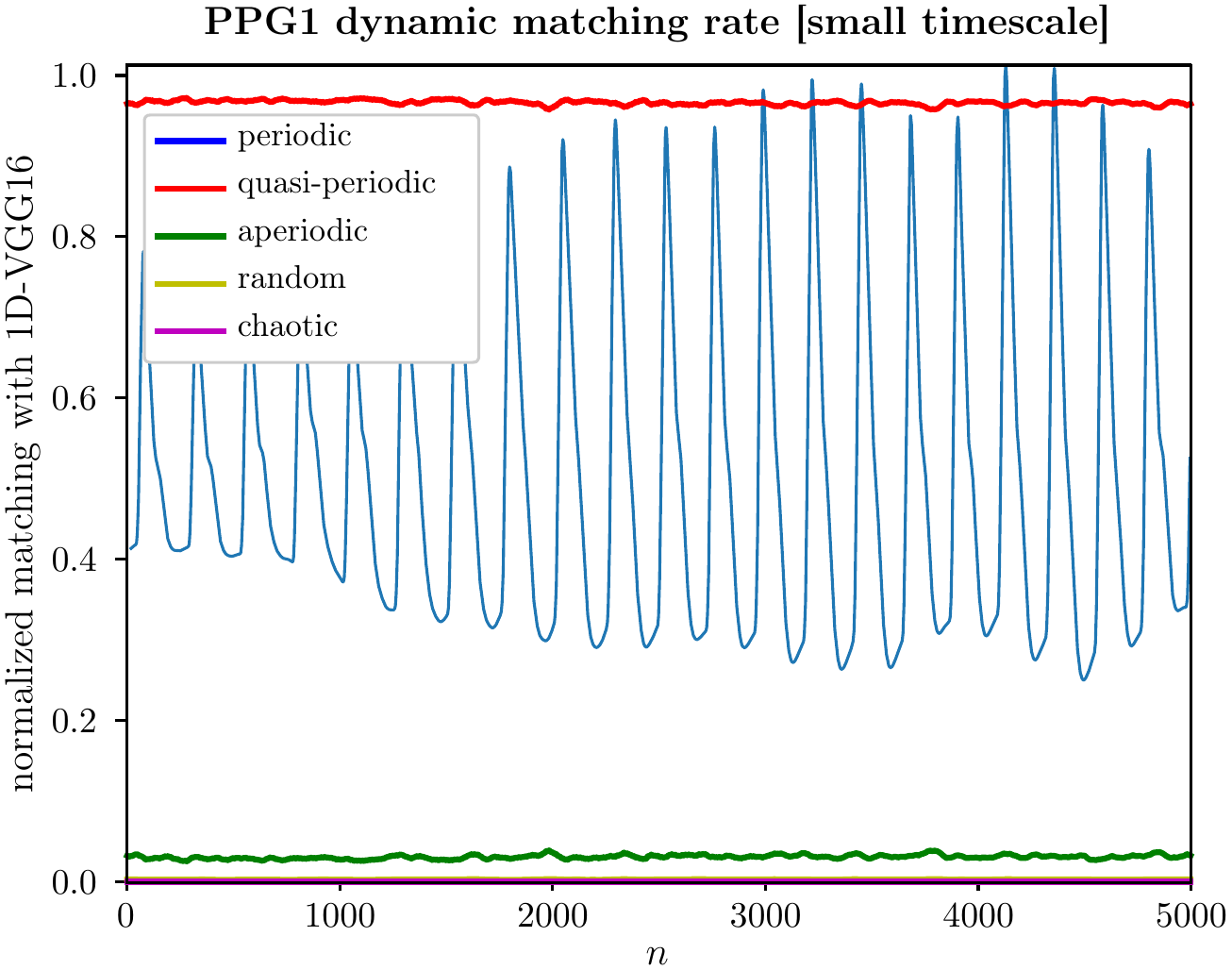}
\label{fig:FCNNDASTA3}}\hfil
\subfloat[]{\includegraphics[width=0.327\columnwidth]{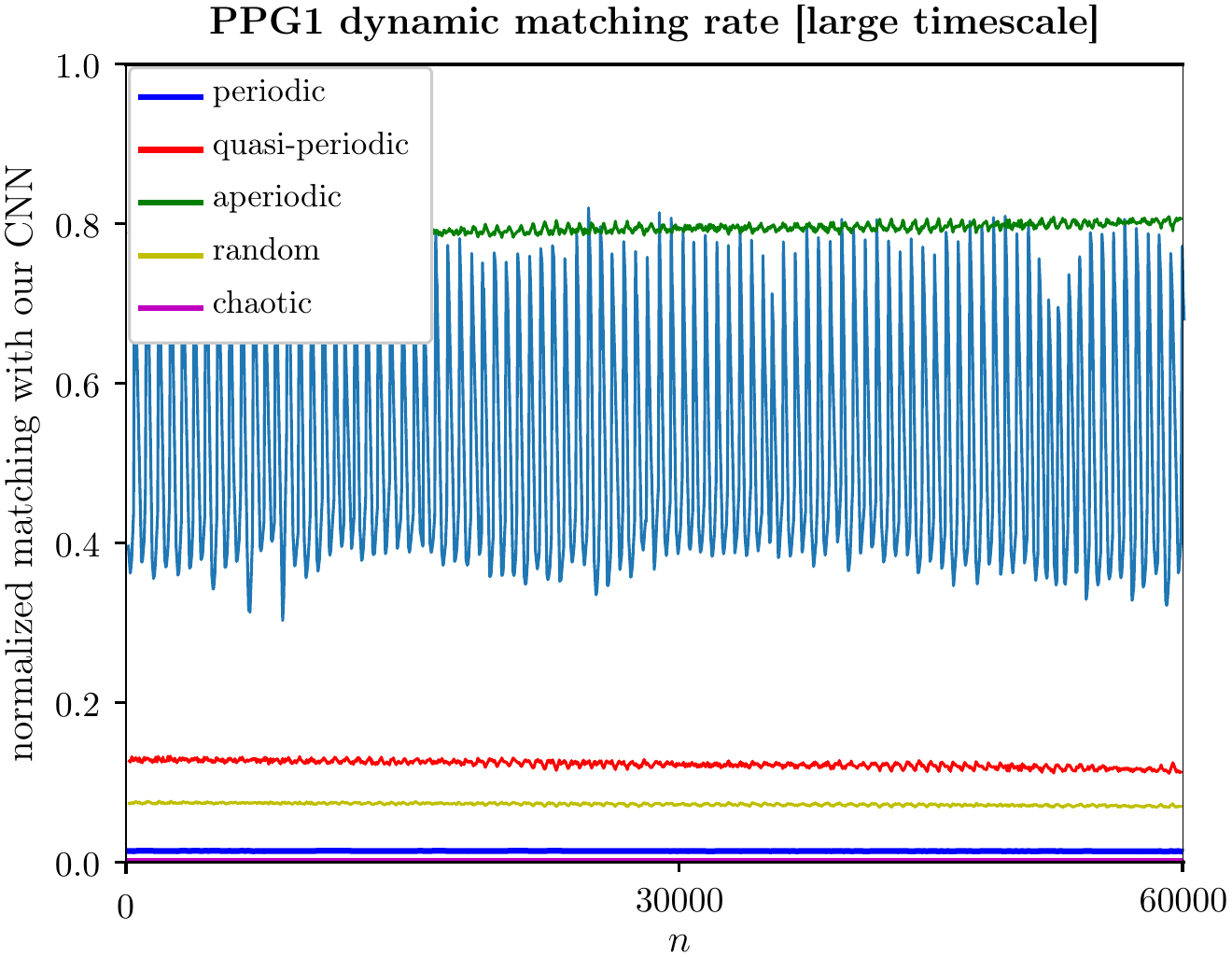}
\label{fig:FCNNDALTA1}}
\subfloat[]{\includegraphics[width=0.33\columnwidth]{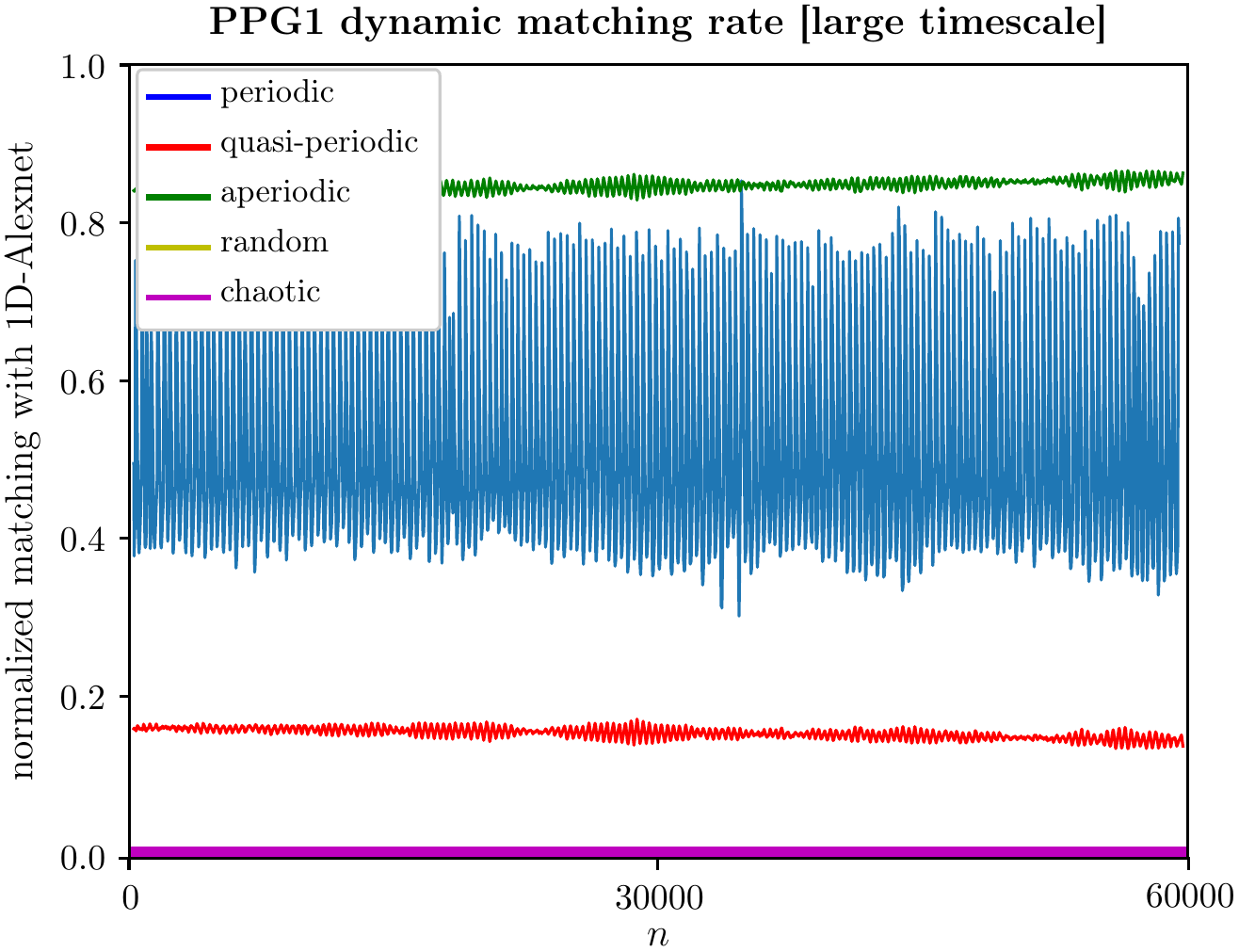}
\label{fig:FCNNDALTA2}}
\subfloat[]{\includegraphics[width=0.33\columnwidth]{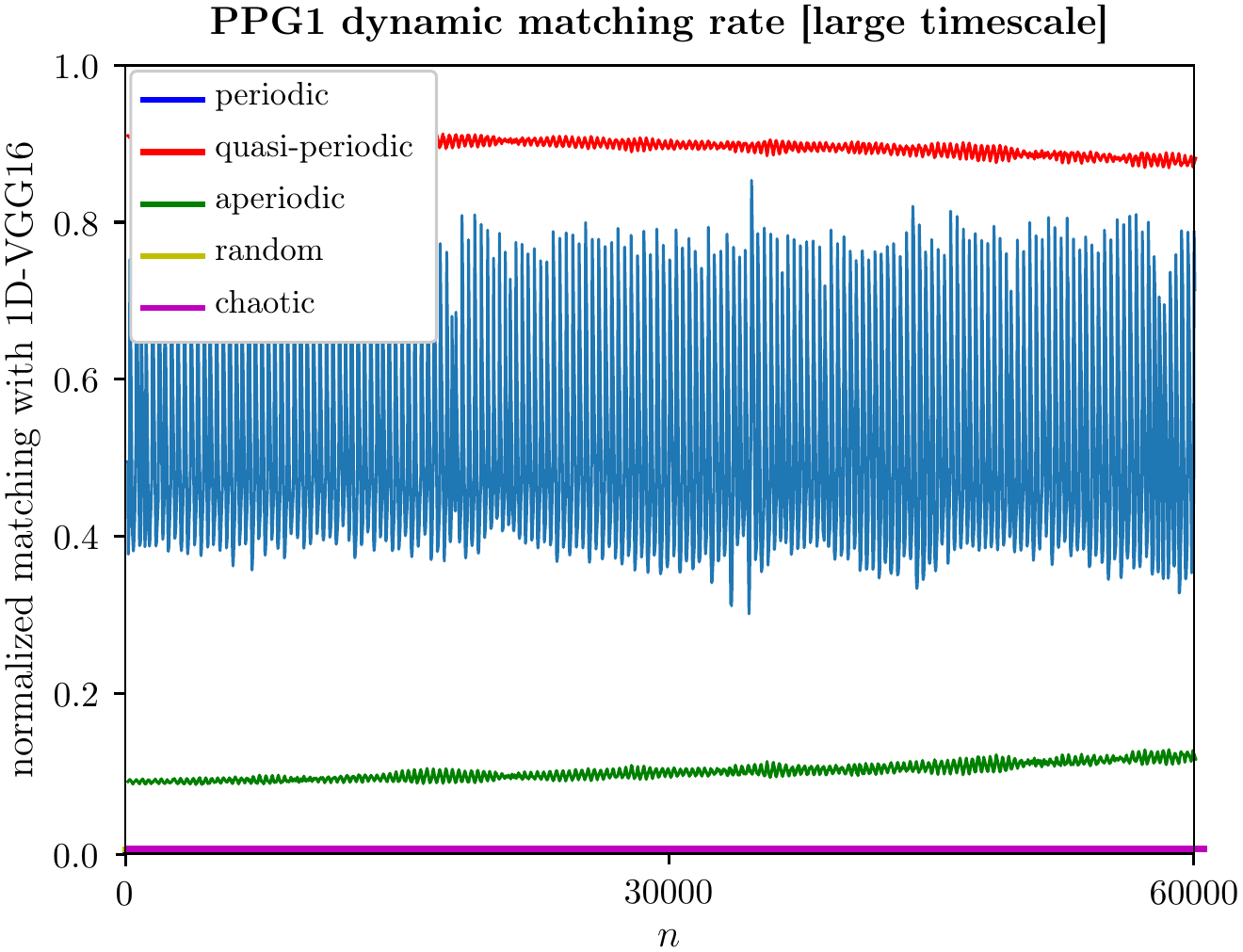}
\label{fig:FCNNDALTA3}}
\caption{From a sample PPG signal (subject number 1): (\textbf{a}) Dynamic behavior on a small timescale with our proposed CNN architecture; (\textbf{b}) Dynamic behavior on a small timescale with the 1D-Alexnet architecture; (\textbf{c}) Dynamic behavior on a small timescale with the 1D-VGG16 architecture; (\textbf{d}) Dynamic behavior on a large timescale with our proposed CNN architecture; (\textbf{e}) Dynamic behavior on a large timescale with the 1D-Alexnet architecture; (\textbf{f}) Dynamic behavior on a large timescale with the 1D-VGG16 architecture.}
\label{fig:FCNNDA}
\end{figure}

\subsubsection{Comparison with other architectures (state of the art)}\label{sssec:comparison}

Finally, in an additional experimental test, we compare our proposed CNN architecture with others from the state-of-the-art to ensure the reliability of the obtained results. The architectures being employed by comparative purposes include Alexnet \cite{Deng2009} and VGG16 \cite{Simonyan2015}, which we have implemented and adapted to solve one-dimensional problems, as the one raises here. The rationale for using these specific state-of-the-art architectures lies in the dissimilarity of their internal structures, each with innovative strategies in information processing.

Starting with the Alexnet architecture, it represents one of the most basic CNN structures. Its conceptualization is appealing because it uses convolutional layers with large kernels and reduces fast these kernels to the small ones, producing abrupt transitions and obtaining more general features of the inputs due to this initial large receptive field of the convolutional layers. As opposed to the Alexnet architecture, we find that the VGG16 architecture introduces another interesting structure in a CNN: the filtering banks. These consist of $3\times 3$ convolutional layers followed by Max Poolings to obtain the greater activations of the feature maps produced. In this case, the transitions between layers are smoother but have a less receptive field, which implies the use of more local features. Lastly, we propose our architecture, which is based on \cite{He2016} and introduces the concept of the residual layers, that allow to recover and use the previous information along the Residual Blocks to avoid the leak of information along the convolutional layers and ensure the use of all the possible information in the CNN to predict the underlying dynamic behavior. As you can see, each of the architectures explained develops important concepts and different types of classification backends, to ensure the reliability of the obtained results.

\begin{table}[ht!]
	\centering
	\caption{Comparative analysis of different CNN-based architectures for four different PPG signals randomly chosen.}\label{tab:CNNDAresults}
	\begin{adjustbox}{max width=\textwidth}
		\begin{tabular}{| c | l | L | L | L | L | L | L |}
			\hline 
			\multirow{2}{*}{\textbf{Signal}} & \multicolumn{1}{c|}{\textbf{Timescale}} & \multicolumn{3}{c|}{\textbf{Small timescales}} & \multicolumn{3}{c|}{\textbf{Large timescales}}\\
			\cline{2-8} 
			& \multicolumn{1}{c|}{dynamics contribution (\%)} & \multicolumn{1}{c|}{1D-Alexnet} & \multicolumn{1}{c|}{Our CNN} & \multicolumn{1}{c|}{1D-VGG16} & \multicolumn{1}{c|}{1D-Alexnet} & \multicolumn{1}{c|}{Our CNN} & \multicolumn{1}{c|}{1D-VGG16}\\
			\hline 
\multirow{5}{*}{PPG1} & periodic component & 0.01  &  0.01 & 0.00  & 0.00  & 1.26  & 0.00 \\
			\cline{2-8} 
			& quasi-periodic component &  98.01 &  99.88 & 94.24  & 15.17  & 11.43  & 89.96 \\
			\cline{2-8} 
			& aperiodic component & 1.96  & 0.05  &  5.35 & 84.82  & 80.51  & 9.88 \\
			\cline{2-8} 
			& chaotic component & 0.00  & 0.03  & 0.00  & 0.00 & 0.02   &  0.00\\
			\cline{2-8} 
			& random component &  0.00 & 0.03  &  0.40 &  0.00 &  6.78 & 0.15 \\
			\hline 
			\multirow{5}{*}{PPG2} & periodic component &  0.01 & 0.01  & 0.00  & 0.00  & 1.24  & 0.00 \\
			\cline{2-8} 
			& quasi-periodic component &  98.11 &  99.89  &  94.18 & 15.58  & 11.04  & 90.28 \\
			\cline{2-8} 
			& aperiodic component & 1.87  & 0.04  & 5.42 & 84.41  &  81.05 &  9.57 \\
			\cline{2-8} 
			& chaotic component &  0.00 & 0.02  & 0.00  & 0.00  &  0.02 &0.00 \\
			\cline{2-8} 
			& random component & 0.00  &  0.03 &  0.39 &  0.00 &   6.64 & 0.14 \\
			\hline 
			\multirow{5}{*}{PPG3} & periodic component &  0.00 & 0.01  & 0.00  & 0.00  & 1.23  & 0.00 \\
			\cline{2-8} 
			& quasi-periodic component &  98.49 &  99.87  &  87.24 &  15.03 &  10.51 &  89.24\\
			\cline{2-8} 
			& aperiodic component & 1.49  &  0.04 & 12.12  & 84.95 & 81.79  & 10.60 \\
			\cline{2-8} 
			& chaotic component & 0.00  &  0.03 & 0.00  & 0.00 & 0.02  & 0.00 \\
			\cline{2-8} 
			& random component & 0.00  & 0.03  &  0.63 & 0.00  & 10.11 & 0.15 \\
			\hline   
			\multirow{5}{*}{PPG4} & periodic component &  0.00 & 0.01  &  0.00 & 0.00  & 1.28  & 0.00 \\
			\cline{2-8} 
			& quasi-periodic component & 98.98 &  99.88  & 99.87  & 22.03  & 23.11  & 97.14 \\
			\cline{2-8} 
			& aperiodic component & 0.99  &  0.05 & 0.10  &  77.95  &  65.47 &  2.78\\
			\cline{2-8} 
			& chaotic component &  0.00 & 0.02 & 0.00  &  0.00 & 0.02  & 0.00 \\
			\cline{2-8} 
			& random component & 0.00 & 0.04  &  0.02 & 0.00  &  10.35 & 0.06 \\
			\hline   
		\end{tabular}
	\end{adjustbox}
\end{table}

In Figure \ref{fig:FCNNDA}, we show a sample input PPG signal superposed with the dynamic behaviors matching-up rate allocated by the proposed CNN-based models at different timescales. In Table \ref{tab:CNNDAresults}, we refer the corresponding numerical values for four different PPG signals randomly chosen. At first sight, according to Table \ref{tab:CNNDAresults}, the three CNN-based architectures reveal that, at small timescales, the predominant dynamic behavior is the quasi-periodic, but at large timescales, the 1D-VGG16 diverges from the norm and puts forward differentiated dynamic behaviors. This discrepancy is linked to the small size of the filters used by the 1D-VGG16 architecture, which improves a local timescale analysis, at the expense of a more global analysis (large timescale). A closer look at the numerical values shows how our CNN model performs a much finer dynamic discrimination than the other two proposed architectures, allowing us to a better identification of a component that we feel could be key in the dynamics of PPG signal, the random component. 

\subsection{Horizon of prediction with an RNN model}\label{sssec:horizon}

\begin{figure}[ht!]
\centering
\includegraphics[width=0.64\columnwidth]{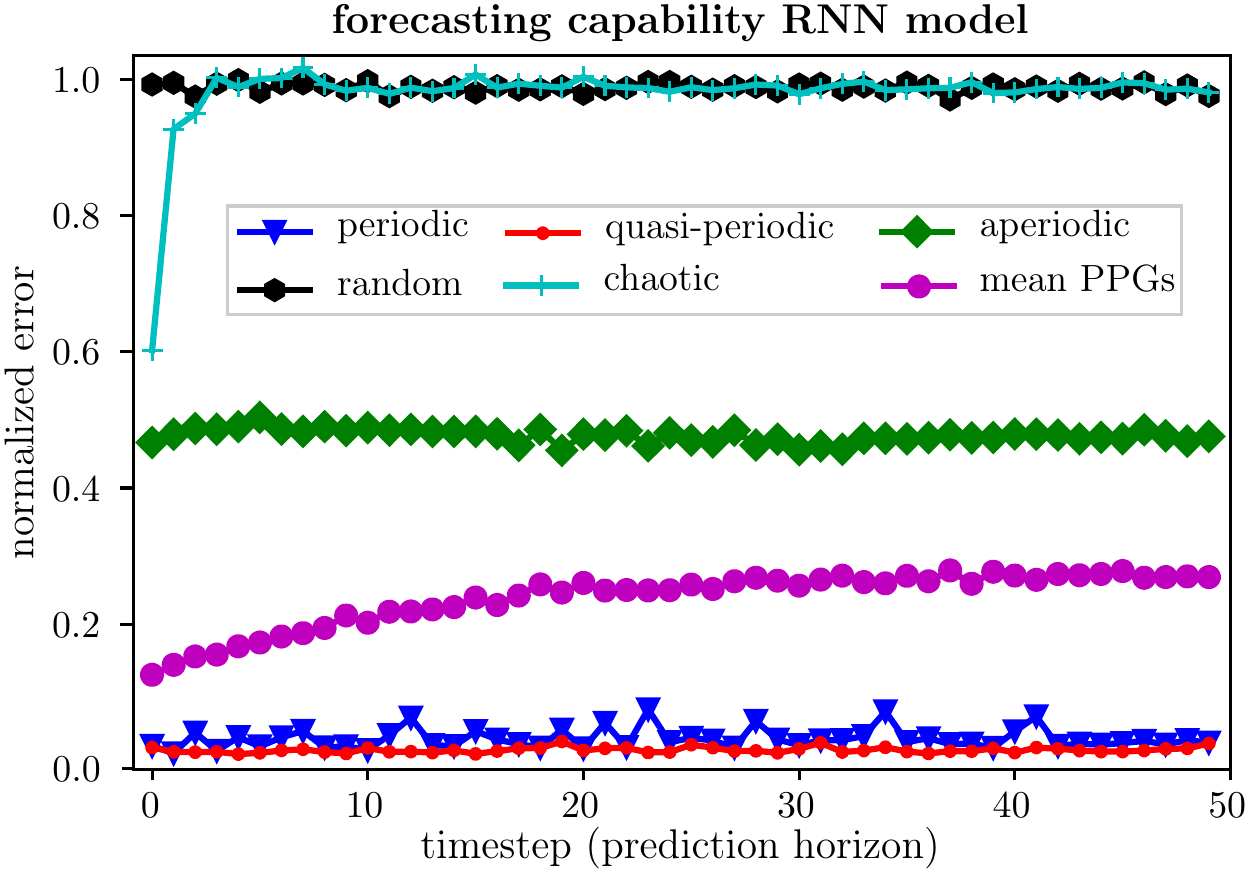}
\caption{Normalized prediction error of detected dynamic behaviors and the average value of five different PPG signals for a horizon of prediction of up to 50 samples.}\label{fig:FHorizonP}
\end{figure}

From another perspective, a supplemental experimental test allows us to predict up to $h_{p}$ future samples of the PPG signal from the current time $t_{c}$. The principal aim of this experimental test is to shed light on how predictable are PPG signal and what type of dynamic behavior fits better in terms of predictability based on the reference signals. With this in mind, we have implemented and adapted a RNN architecture (cf. \S\ \ref{ssec:RNNarch} for more details) to infer a nonlinear regression model that best fits both PPG and reference signals. All the results are shown in Figure \ref{fig:FHorizonP}.

\section{Discussion}\label{sec:Discussion}

We tackle the dynamical analysis of the PPG signal of young and healthy individuals from two complementary angles. On one side, we identify the dynamic patterns of behavior present in the PPG signal at different timescales with the aid of a custom CNN architecture. On the other, we consider the level of predictability of the PPG signal on the grounds of a nonlinear regression model implemented in an RNN architecture.

\subsection{Dynamic behavior classification}

Within physiological systems, the seeming regularity at a glance of the time evolution of many biological signals, as is the case with the PPG signal, covers up fairly more complex dynamic patterns, slow-onset processes, that contribute to efficiently regulating homeostatic mechanisms to the body to function properly. The proposed CNN model has allowed us to classify the dynamics of the PPG signal of young and healthy individuals at different timescales, in connection with the most typical dynamics found on a wide range of physical systems, i.e., periodic, quasi-periodic, aperiodic, chaotic or random dynamic behaviors.

In the first instance, we analyze the dynamics of all PPG signals available in an experimental project conducted in 2015 by using the dynamic classification process implemented in the CNN model. The number of participants in this research project amounted to 40 students between 18 and 30 years old and non-regular consumers of psychotropic substances, alcohol or tobacco. According to Table \ref{tab:CNNresults} (cf. \S\ \ref{sssec:ourCNN}), the dynamic behavior of the PPG signal is clear and predominantly quasi-periodic at small timescales, in line with the physiological response of the cardiorespiratory system, which manifests in the form of a pulsatile component, due to the heartbeat, modulated quasi-periodically by breathing pattern, among other factors. However, on a large timescale, the PPG signal starts experimenting with a strong tendency to aperiodic behavior, despite the presence of some latent quasi-periodicity, which sometimes fades or becomes confused with a less remarkable periodical bias. The cardiorespiratory system as a whole is affected by physiological processes that develop slowly, to better aligning to body's needs in all time, which justifies a more competition of factors affecting cardiac modulation. In either case, in the analysed PPG signals, we have found no significant trace of chaotic or random behavior at all different timescales.

 To guarantee obtained results, we also evaluate the influence of filtering and noise on the dynamics of PPG signals. So, Table \ref{tab:CNNRFresults} (cf. \S\ \ref{sssec:filter}) shows the average contribution that the CNN model assigns to the dynamics present in all PPG signals at different timescales, before and after filtering. As we can see, the raw PPG signals available to us in the study have quite a lot of motion artifacts or artifacts that arose out of the conditions in which the measurements were conducted. As a result, our CNN model has difficulties in clearly distinguishing between the two predominant dynamic behaviors at different timescales, quasi-periodic and aperiodic dynamics, emerging sudden fluctuations that confuse and deceive the matching-up process of the CNN model, as exemplified in \Crefrange{fig:FCNNRST}{fig:FCNNRLT}. However, once PPG signals are filtered with a simple Butterworth bandpass filter with cutoff frequencies at 0.01 and 8 Hz---in order to avoid high-frequency noise and to some extent, motion artifacts, without losing as much dynamic information as possible---, the matching-up process of the CNN model does not show fluctuations and provides a clear dynamic response. So, on a small timescale a quasi-periodic dynamic behavior and on a large timescale an aperiodic behavior, as shown as examples in \Crefrange{fig:FCNNFST}{fig:FCNNFLT}, respectively. Nevertheless, despite the predominant aperiodic behavior, our CNN model recognises two additional underlying components: a quasi-periodic component and a random component. One of them, the first, reveals an about sustained heart rate, quasi-periodically modulated by breathing rhythm under normal operating conditions. The second of these, we think that provide the stochastic component that breaks up the apparent dynamic inflexibility and is responsible for introducing multiple frequency variations much needed in a system as adaptive as the cardiorespiratory system (see Figure \ref{fig:FCNNP1BLT} for an example).

In the case of the effect of noise on the dynamics of PPG signals, we introduce additional levels of white noise amplitude (1\%, 5\% and 10\%) to the PPG signals to verify the dynamic tolerance of PPG signals to external noise. As can be seen on Table \ref{tab:CNNWNresults} (cf. \S\ \ref{sssec:noise}), up to about 5\% our CNN model is being able to retain the prevailing deterministic dynamic response at different timescales. Simply put, on a small timescale a quasi-periodic dynamic behavior, and on a large timescale an aperiodic behavior, although now with 5\%, the presence of a strong random component in the detected dynamic behavior is noticeable; with 10\% additive noise, the predominant dynamic behavior is random (see Figure \ref{fig:FCNNWN} for an example).        

Finally, we compare the performance of the classification process of our CNN model with that of other implementations of remarkable CNN architectures. In light of the results of Table \ref{tab:CNNDAresults}, we can reaffirm that our proposed architecture performs a dynamical analysis very consistent with other typical configurations based on convolutional neural networks. Our proposed model and the Alexnet model use wider-band filters, enabling the detection of a very low frequency dynamic spectrum (large timescale), unlike the 1D-VGG16 model, where the small size of the filters make correct detection of the dynamics on a large timescale virtually impossible. A more thorough examination of the numerical data show how our CNN architecture becomes significantly more selective, in terms of dynamic discrimination, than the other two proposed architectures, allowing us to a better identification of, among others, the random component, which we see as a key factor in the dynamics of PPG signal at large timescales.  

\subsection{Horizon of prediction}

From another angle, since we seek to show if the PPG signal of young and healthy individuals is chaotic, we proceed to evaluate the predictive capacity of an RNN model, which implements a nonlinear regressive model, both on PPG signals and reference signals. It is well known that random systems are unpredictable and, therefore, their prediction error is maximal for any prediction horizon. Chaotic systems become unpredictable in a very short space of time, because initially, very close trajectories soon diverge exponentially. Periodic and quasi-periodic signals are completely predictable and, hence, their prediction horizon is virtually limitless, or, in other words, their prediction error is almost zero. Aperiodic signals are somewhat predictable, as initially very close trajectories either evolve in parallel or diverge linearly over time. In any case, the prediction error is always bounded. 

Figure \ref{fig:FHorizonP} (cf. \S\ \ref{sssec:horizon}) shows the prediction horizon of all signals studied, with an error normalized between 0 and 1. The predictable behavior of our RNN architecture for all reference signals meets the above criteria. In the case of PPG signals, the predictable behavior of our RNN architecture shows a bounded prediction error, while lower than aperiodic signal and always away from chaotic and random dynamic regimes, as we can see in Figure \ref{fig:FHorizonP}. We think that the PPG signal is an intricate combination of two different main dynamics, a regular rhythm imposed by the heart rate and an underlying aperiodic dynamic that develops very slowly. Accordingly, there is an excellent balance between all of them,          depending on the physiological requirements of each individual. Hence, in the short term, at the beginning of the prediction horizon, a predictive behavior closer to the typical quasi-periodicity prevails, but as the length of the prediction horizon increases, the aperiodic behavior comes in more precisely. So, an increasing dynamic complexity leads to predictive behavior with a growing but bounded prediction error. It is within the realms of possibility that with an infinite predictive horizon the prediction error will increase very gradually, as more slowly developing physiological factors take part in the overall dynamics (stronger aperiodicity), but under no circumstances, would it become unpredictable, as it would in chaotic or random dynamic behaviors.   

\section{Conclusions}\label{sec:Conclusions}

In this paper, we show, through different experimental tests based on neural networks, that the PPG signal dynamics of young and healthy individuals are not chaotic or random on any timescale. Nevertheless, behind the apparent regularity of a PPG signal, there is a hidden increasing dynamic complexity with timescale. On a small timescale, the dominant dynamics of a PPG signal is mainly attributable to a quasi-periodic behavior, but on a large timescale, the dynamics of a PPG signal become aperiodic, although it does not become chaotic or random. It has been shown both with a classification tool of the present dynamics in a PPG signal, at different timescales, and with the predictive capability of a nonlinear regression model.

We think, pending confirmation by ongoing research work---while preliminary results are steps in the right direction---, that the aperiodic dynamics of a PPG is consistent with the characteristic behavior of a strange nonchaotic attractor (SNA). This category of attractors acts as a dynamic interface between quasi-periodic and chaotic behavior. As an active agent that a SNA is in the transition towards chaos, a highly promising line of research involves finding out how the stochastic (random) component present on a large timescale in a PPG signal facilitates the transition, with all of the consequences that that entails, as well as the psychosomatic causes, or even chronic physiological conditions, that could affect its development.

\section*{Acknowledgments}
The authors would like to thank Life Supporting Technologies Group (LST-UPM) for taking part in project FIS-PI12/00514, from MINECO. Also, they want to thank Google for the opportunity to use their Google Collaboratory services for free, which have helped to make parts of this work possible.

\section*{Abbreviations}

The following abbreviations are used in this manuscript:\\

\noindent 
\begin{tabular}{@{}ll}
ADAM & Adaptive Moment Estimation \\
CNN & Convolutional Neural Network \\
DNN & Deep Neural Network \\
LSTM & Long Short-Term Memory \\
MDPI & Multidisciplinary Digital Publishing Institute \\
PPG & PhotoPlethysmoGraphy \\
RNN & Recurrent Neural Network \\
SGD & Stochastic Gradient Descent \\
\end{tabular}

%\section*{References}

\bibliographystyle{apalike}
\bibliography{Paper3arXiv}

\end{document}